\documentclass[useAMS,usenatbib]{mn2e}
\usepackage{graphicx}
\usepackage{times}
\usepackage{colordvi}
\usepackage{epsfig}
\begin{document}
\title[Abundances in Low-redshift DLA systems]{Elemental Abundance 
Measurements in Low-redshift Damped Lyman-$\alpha$ Absorbers }

\author[J. D. Meiring, V. P. Kulkarni et al.]{Joseph D. Meiring$^{1}$, 
Varsha P. Kulkarni$^{1}$, Pushpa
Khare$^{2}$, Jill Bechtold$^{3}$,  
Donald G. York$^{4,5}$,  
\newauthor Jun Cui$^{3}$, James T. Lauroesch$^{6}$, Arlin P. S. Crotts$^{7}$, Osamu Nakamura$^{8}$ \\ 
$^{1}$Department of Physics and
Astronomy, University of South Carolina, Columbia, SC 29208, USA \\
$^{2}$Department of Physics, Utkal University, Bhubaneswar, 751004, India \\
$^{3}$Department of Astronomy and Astrophysics, University of Arizona, Tucson, AZ 85721, USA \\ 
$^{4}$Department of Astronomy and Astrophysics, University of Chicago, Chicago, IL 60637, USA \\ 
$^{5}$Enrico Fermi Institute, University of Chicago, Chicago, IL 60637, USA \\ 
$^{6}$Department of Physics and Astronomy, Northwestern University, Evanston, IL 60208, USA \\ 
$^{7}$Department of Astronomy, Columbia University, New York, NY 10027, USA\\ 
$^{8}$School of Physics and Astronomy, University of Nottingham, NG7 2RD, UK}


\date{Accepted ... Received ...; in original form ...}

\pagerange{\pageref{firstpage}--\pageref{lastpage}} \pubyear{}

\maketitle

\label{firstpage}

\begin{abstract}
 We present elemental abundance measurements for 9 damped Ly-$\alpha$ systems 
 (DLAs) and 1 sub-DLA 
 at 0.1 $\la$ z $\la$ 1.5 from recent observations with the Multiple Mirror Telescope.
 Most of these absorbers are found to be metal-poor, while 2 are found to have 
 $\approx 30-50 \%$ solar metallicities. Combining our data with other data from the literature, we find that 
 the systems with higher [Zn/H] also have stronger depletion as measured by [Cr/Zn] and [Fe/Zn]. 
 The relationship between the metallicity and H I column density is also investigated. Together with our previous MMT survey
  (Khare et al. 2004) we have discovered 2 of the 4 known absorbers at $z < 1.5$ that lie above (although near) the ``obscuration threshold''. 
 This appears to be a result of selecting absorbers with strong metal lines in our sample. 
 It would be interesting to find other similar systems by observing a larger sample and study how much such systems
 contribute to the cosmic budget of metals. Finally, an analysis of the $N_{\rm H I}$-weighted mean metallicity vs. redshift for 
 our sample combined with data from the literature supports previous conclusions 
 that the $N_{\rm H I}$-weighted mean global DLA metallicity rises slowly at best and falls short of solar 
 levels by a factor of $>$ 4 even at $z=0$.
\end{abstract}


\section{Introduction}

Absorption line systems in QSO spectra provide a unique window into the high-redshift universe. These systems have been detected
 in the range 0 $\la z \la 6$, corresponding to $\sim$ 90$\%$ of the age of the universe.  As the absorption line strengths
 depend primarily on the gas content, they are selected independent of the stellar luminosities of the galaxies.
 Thus in principle they may be expected to provide less biased probes of galaxies than objects used
 to study the high redshift universe through emission, which are biased toward brighter galaxies including actively star forming galaxies
 such as Lyman-break galaxies, active galaxy nuclei, and QSOs. 

Damped Lyman-$\alpha$ Absorption systems (DLAs) with log N$_{\rm HI}$ $\ge$ 20.3 and sub-DLAs with 19.0 $\le$ log N$_{\rm HI} <$
 20.3 seen in QSO spectra can be used as probes of the neutral gas content of the universe. DLAs and sub-DLAs are known to
 contain the majority of the neutral gas in the universe (e.g. Wolfe et al. 1995, Peroux et al. 2003). With such high neutral
 hydrogen column densities DLAs are believed to be closely linked to galaxies. DLAs may also be the neutral gas reservoirs used in
 star formation. Furthermore, DLAs and sub-DLAs provide detailed gas-phase abundances of a number of elements and can thus be used
 to test galactic chemical evolution models.

Several elements have been detected in DLAs including C, N, O, Mg, Si, S, Ca, Ti, Cr, Mn, Fe, Ni, and Zn. 
 Zn is often used as a tracer of gas-phase metallicity as it is typically relatively undepleted in the Galactic ISM when the
 fraction of H in molecular form is low, which is the case for DLAs \citep{Sem95}, Zn tracks the abundance of Fe in
 Galactic stars \citep{Mis02}, and the lines of Zn II $\lambda$$\lambda$ 2026,2062 are fairly weak
 and unsaturated. Also, Zn II can be detected with ground-based observations over a wide range of redshifts (0.6 $\la z \la 3.
5$).  Elemental abundances relative to Zn such as [Cr/Zn] or [Fe/Zn] can be used to estimate the dust content of the QSO absorber.
Here and throughout the paper we adopt the standard notation [X/H] = log (N$_{\rm X}$/N$_{\rm H I}$) - log (X/H)$_{\sun}$.  

If DLAs do constitute a representative sample of galaxies, it would be expected that their global mean metallicity of these objects
 should rise to roughly solar values at low redshift. It is uncertain whether or not DLAs show this
 trend. The most recent evaluations of this by Kulkarni and Fall (2002), Prochaska et al. (2003b), and Kulkarni et al. (2005) show
 at best a weak evolution in the metallicity-redshift relation with slopes of $\sim$ - 0.20$\pm$0.07. 

One proposed explanation of the lack of metal-rich DLAs in observed samples is that this population is being under-sampled
 due to dust obscuration (e.g.; Boisse et al. 1998; Pei $\&$ Fall 1993; Vladilo $\&$ Peroux 2005).
 Absorbers with higher metallicity may also contain more dust, making any background QSOs fainter and keeping 
 those objects out of magnitude-limited samples \citep{York06}. In fact, it has been claimed that up to 50 $\%$ of QSOs may be hidden
in optical surveys due to dust obscuration \citep{Vlad05}. Recently, Akerman et al. (2005) measured the metal abundances of $z > 1.8$ 
 DLAs from the Complete Optical and Radio Absorption Line System survey (CORALS), and found that N$_{\rm HI}$ weighted mean metallicity
 based on the Zn abundance is only marginally higher but within error limits of previous samples. However, the effects of dust may be expected
 to be stronger at low redshifts.  The role of dust in DLAs is still unclear, and more investigation is needed. We discuss dust and
 reddening in our sample of absorbers in more detail in $\S$ 5.2.

Assuming a cosmology with $\Omega_{m}$ = 0.3 and $\Omega_{\Lambda}$ = 0.7, $z < $ 1.5 spans $\sim$70
 $\%$ of the age of the universe. It is therefore critical to obtain more abundance measurements at low
 redshift. One reason for the uncertainty in the metallicity-redshift relation
 is the lack of abundance measurements in low-redshift absorbers. In fact $\sim$ 70$\%$ of the existing Zn measurements are for
 2 $\la$ z $\la$ 3.5, which corresponds to only $\sim$ 10$\%$ of the age of the universe. The paucity of Zn measurements at
 low  z is primarily because the Zn $\lambda$$\lambda$ 2026, 2062 lines lie in the UV for z $<$ 0.6, and deeply in the blue end of the
 spectrum for 0.6 $<$ z $\la$ 1.3, where most existing spectrographs are inefficient. In this paper we present new MMT observations
 toward 10 QSOs. The observations presented in this paper increases the
 size of the sample of absorbers at z  $\la$ 1.5 by $\sim$ 40$\%$.

 In $\S$ 2 we describe our observations and our data reduction routines. $\S$ 3 describes the estimation of column densities. In
 $\S$ 4 we present notes on the individual objects. We present results from these measurements in $\S$ 5, and in $\S$ 6 we
 discuss our results and their implications. 
 
\section{Observations and Data Reduction}
  Our sample consists of 10 confirmed DLA and sub-DLA absorbers at 0.2 $\la$ $z$ $\la$ 1.5, for which N$_{\rm H I}$ is known
 from $HST$ spectra \citep{Rao05}. Seven of these targets were observed in the Sloan Digital Sky Survey (SDSS),
 and most show the presence of strong Mg II or Fe II absorption features from SDSS spectra.  Apparent magnitudes for these 7 QSOs are
 available from the SDSS Quasar Catalog III, Data release 3 \citep{Abaz05}.

 The spectra presented here were obtained at the Multiple Mirror Telescope (MMT) during two separate epochs, 2004 May and 2005
 February. Nearly 2.7 nights out of a total of 5 were lost due to poor weather. The Blue Chanel spectrograph was used with the 832 l 
 mm$^{-1}$ grating in the first order, or the 1200 l mm$^{-1}$ grating in the second order, depending on the redshift of the absorber. 
 A CuSO$_{4}$ blocking filter was used with the 832 l mm$^{-1}$ grating to block first order red light. The central wavelength for the
 832 l mm$^{-1}$ grating
 was at 3670 $\mbox{\AA}$ and 3600 $\mbox{\AA}$ for the 2004  May and 2005 Feb epochs respectively. For the 1200 l mm$^{-1}$ grating 
 the wavelength was centered at 4935 $\mbox{\AA}$
 and 4770  $\mbox{\AA}$ during the 2004  May and 2005 Feb epochs respectively. In order to achieve the desired S/N while minimizing
 cosmic rays, multiple exposures of each target were taken with exposure times ranging from 1800 to 2700 s depending on the magnitude of
 the QSO. Each target observation was followed by a comparison spectrum from a He+Ne+Ar lamp for wavelength calibration. Quartz flat
 fields and bias frames were taken at both the beginning and end of the night.  Table 1 lists the
 vacuum wavelengths and oscillator strengths used in identification of the features and subsequent analysis.Table 2 gives a summary of our
 observations.

\begin{table}
\begin{center}
\caption{Atomic Data}
\begin{tabular}{lccc}
\hline
\hline
Species & $\lambda_{rest}$ & $f$ & References \\
        & $\mbox{\AA}$    &     &                      \\
\hline
Be II       &   3130.4219 & 3.321E-1 &  1   \\
C IV        &   1548.2040 & 1.899E-1 &  1   \\
            &   1550.7810 & 9.475E-2 &  1    \\ 
Mg I        &   1827.9351 & 2.420E-2 & 1  \\
	    &   2026.4768 & 1.130E-1 & 1 \\
	    &   2852.9631 & 1.830E00 & 1  \\
Mg II       &   2796.3543 & 6.155E-1 & 1 \\
            &   2803.5315 & 3.058E-1 & 1 \\
Al II       &   1670.7886 & 1.710E00 & 1 \\
Al III      &   1854.7184 & 5.590E-1 &  1 \\
            &   1862.7910 & 2.780E-1 &  1 \\
Si II       &   1808.0129 & 2.080E-3 & 1,2 \\
Si IV       &   1393.7602 & 5.130E-1 & 1    \\
            &   1402.7729 & 2.540E-1 &  1   \\
Ca II       &   3933.6614 & 6.267E-1 & 1   \\
            &   3968.4673 & 3.116E-1 &  1   \\
Ti II       &   1910.6123 & 1.040E-1 & 1,3 \\ 
            &   3383.7588 & 3.580E-1 & 1  \\
Cr II       &   2056.2569 & 1.030E-1 &  1,2 \\
            &   2062.2361 & 7.590E-2 & 1,2 \\
Mn II       &   2576.8770 & 3.610E-1 &  1  \\
            &   2594.4990 & 2.800E-1 &  1   \\
            &   2606.4620 & 1.980E-1 & 1   \\
Fe II       &   2249.8768 & 1.820E-3 &  4  \\
            &   2260.7805 & 2.440E-3 & 4   \\
            &   2344.2139 & 1.140E-1 & 1,5 \\
            &   2374.4612 & 3.130E-2 &  1  \\
            &   2382.7652 & 3.200E-1 & 1,5 \\
Zn II       &   2026.1370 & 5.010E-1 & 1,6 \\
            &   2062.6604 & 2.460E-1 & 1,6 \\
\hline
\end{tabular}
\vspace{0.2cm}
\begin{minipage}{140mm}
References. (1) Morton 2003; (2) Bergeson $\&$ Lawler 1993b; \\ (3) Wiesse et al. 1996;  
                 (4) Bergeson et al. 1994; (5) Bergeson et al. 1996;  \\ (6) Bergeson $\&$ Lawler 1993a
\end{minipage}
\end{center}
\end{table}

\begin{table*}
{\footnotesize
\caption{Summary of Observations}
\begin{tabular}{ccccccccc}
\hline
\hline
QSO & $g$ & $z_{em}$ & $z_{DLA}$ & log N$_{\rm H I}$ & Grating & FWHM  & Exposure Time & Epoch  \\
    &     &          &           &     cm$^{-2}$      &   lines mm$^{-1}$ & $\mbox{\AA}$ &        &         \\
\hline
0738+313  & 16.1 & 0.630        & 0.0912  & 21.18$\pm$0.06   & 832   & 1.10 & 3000 & 2005 Feb.       \\
$\cdots$  & $\cdots$ & $\cdots$ & 0.2210  & 20.90$\pm$0.09 & 832   & 1.10 & 3000 & 2005 Feb.         \\
0827+243  & 17.3 & 0.941        & 0.5247  & 20.30$\pm$0.05 & 832   & 1.10 & 3600 & 2005 Feb.         \\
1010+0003 & 18.3 & 1.398        & 1.2651 & 21.52$\pm$0.07  & 1200  & 1.33 & 3600  & 2005 Feb.        \\
1107+0003 & 18.7 & 1.741        & 0.9547  & 20.26$\pm$0.14 & 832   & 1.10 & 8100 & 2005 Feb.         \\   
1137+3907 & 17.4 & 1.023        & 0.7190  & 21.10$\pm$0.10 & 832   & 1.10 & 5400 & 2004 May         \\
1225+0035 & 18.9 & 1.226        & 0.7731  & 21.38$\pm$0.11 & 832   & 1.10 & 5400 & 2005 Feb.         \\
1501+0019 & 18.1 & 1.930        & 1.4832 & 20.85$\pm$0.13 & 1200  & 1.33 & 5400  & 2004 May         \\
1712+5559 & 18.7 & 1.356        & 1.2095 & 20.72$\pm$0.05 & 1200  & 1.33 & 5400 &  2004 May         \\
1715+4606 & 18.1 & 0.989        & 0.6511  & 20.44$\pm$0.13 & 832   & 1.10 & 5400 & 2004 May         \\
1733+5533 & 18.1 & 1.074        & 0.9981  & 20.70$\pm$0.03 & 832   & 1.10 & 3600 & 2004 May         \\

\hline
\end{tabular}
}
\end{table*}

Data reduction was carried out using standard IRAF routines. Images were bias subtracted, trimmed, and flat fielded using the task
 CCDPROC. These reduced 2-d spectra were then extracted and wavelength calibrated using the task DOSLIT. The wavelengths were
 converted to vacuum using the task DISPTRANS. Finally the spectrum was normalized using the  task CONINUUM. The solar abundances 
used as reference while calculating the abundances are adopted from Lodders et al. (2003).

Figures 1-10 plot the co-added and continuum-fitted spectra obtained with the MMT, along with line identifications from the DLAs
 as well as other absorption systems along the line of sight. Unmarked, minor depressions have been verified to be artifacts.
 
\section{Estimation of Column Densities} 
 Column densities were estimated by fitting the observed absorption line profiles using the package FITS6P \citep{Wel91}, which has
 evolved from the code by Vidal-Madjar et al (1977). FITS6P iteratively minimizes the $\chi^{2}$ value between the data and
 a theoretical Voigt profile that is convolved with the instrumental profile. All lines were fit with a single component Voigt profile. If
 a multiplet was available such as Fe II $\lambda$$\lambda$ 2344, 2374, 2382 the lines were fit simultaneously until convergence for a
 common N, $b_{eff}$, and $v$. Equivalent width measurements were obtained with the package SPECP, also written by D.E. Welty. 
 The 1$\sigma$ equivalent width uncertainties were estimated using the photon noise uncertainties and the continuum uncertainties obtained
 by allowing the continuum to vary by $\pm$ 10 $\%$. In all cases these uncertainties were dominated by the photon noise uncertainties. 

 The prescription  of Khare et al. (2004) for fitting the Cr II and Zn II lines of was followed. Specifically, 
 the Cr II $\lambda$$\lambda$2056, 2066 lines were fit simultaneously to obtain N, $b_{eff}$, and $v$ values.  Then, the blended Cr II+Zn
 II  $\lambda$ 2062 line was fit starting with the values obtained from the previous fits, holding the Cr II component fixed.
 The Zn II $\lambda$ 2026 line was then fit using the $b_{eff}$ and $v$ value from the Cr $\lambda$$\lambda$ 2056, 2062 fit while holding
 the Mg I component fixed. If the Cr II $\lambda$ 2056 line could not be fit due to noise, we fit the Zn II $\lambda$2026 line, holding
 the Mg I $\lambda$  2026 contribution fixed, then fit the Cr II + Zn II $\lambda$ 2062 blend holding the Zn II component fixed. The Mg I column density was
 estimated by fitting the Mg I $\lambda$ 2852 line from the SDSS spectra. 
 York et al. (2006) showed that the Mg I $\lambda$ 2852 lies on the linear portion of the curve of growth for W$^{\rm rest}_{\rm Mg I}$ $\sim$
 0.6 to 0.7. The Mg I $\lambda$ 2852 line is often moderately saturated in DLAs and the assumption of a linear curve of growth may lead to
 an underestimation of the Mg I column density, therefor we did not assume Mg I $\lambda$ 2852 to be on the linear portion of the curve 
 of growth, and instead used the results of a profile fit to that line. To judge the effects of possible saturation of Mg I $\lambda$ 2852
 on N$_{\rm Mg I}$ and hence on N$_{\rm Zn II}$, we also estimated the maximum N$_{\rm Mg I }$ that would be consistent with W$_{\rm 2852}$
 within a 3$\sigma$ level. We then used this maximum N$_{\rm Mg I}$ to estimate the minimum N$_{\rm Zn II}$ in the $\lambda$ 2026 line.  
  In any case, in our experience the contribution of Mg I to the blended $\lambda$ 2026 line 
 is typically small (eg. Peroux et al. 2006a). For Q1501+0019 we also have a constraint on N$_{\rm Mg I}$ from the non-detection of the Mg I 
 $\lambda$ 1827 line which gives log N$_{\rm Mg I}$ $<$ 13.47.

 Tables 3 and 4 list the derived equivalent widths and column densities respectively. If a certain species was not detected, 
 3$\sigma$ upper limits were placed on the equivalent widths and column densities based on the signal to noise ratio (S/N) in the region 
 based on a 3 pixel resolution element. It was assumed that these lines lie on the linear portion of the curve of growth. As can be seen
 in figures 1-10, none of the detected Zn II or Cr II lines are saturated. In fact, none of the lines detected appear to show significant saturation even in those cases where the
 equivalent widths are large. This is likely due to the line profile being composed of multiple components that could not be resolved at
 the resolution of our data.

\section{Discussion of Individual Objects}

$Q0738+313$ ($z_{em}$ = 0.630, $z_{abs}$ = 0.0912 for system A and $z_{abs}$ = 0.2210 for system B): Both of the systems listed are DLAs.
 The Zn and Cr lines were studied by Kulkarni et al. (2005) using $HST$ data.  
 The $\lambda$$\lambda$ 3933,3969 lines of Ca II appear in both 
 systems A and B.  We also placed upper limits on Ti II $\lambda$ 3383 of system B. As the background QSO is very bright (g = 16.1), we
 were able to obtain high S/N of $\sim$ 100 in the region. This gives an upper limit of log (N$_{\rm Ti II}$/N$_{\rm H I}$) $<$ -9.42,
 with [Ti/H] $<$ -2.32. The feature located at $\sim$ 5070 $\mbox{\AA}$ is an artifact from bad pixels in the CCD. 

$Q0827+243$ ($z_{em}$ = 0.941, $z_{abs}$ = 0.259 for system A and $z_{abs}$ = 0.5247 for system B): The Zn and Cr lines of system B were
 studied in detail at lower wavelengths by Khare at al. (2004). The Fe II $\lambda$$\lambda$ 2249, 2260, 2344, 2374, 2382, 2586, 2600
 lines were measured previously. Strong Mg II $\lambda$$\lambda$ 2796, 2803 and Mg I $\lambda$ 2852 lines were present in system B in the
 new spectra. The Mg I $\lambda$ 2852 line was fit first to estimate $b_{eff}$ and $v$ values, which were used when fitting the Mg II
 doublet. Due to the strong saturation of the Mg II $\lambda$$\lambda$ 2796, 2803 doublet we give the derived column density 
 as a lower limit. We were able to place upper limits on Ti II $\lambda$ 3383 of [Ti/H] = -1.45 at S/N $\sim$ 60 in the region. The
 feature located at $\sim$ 5070 $\mbox{\AA}$ is again an artifact from bad pixels in the CCD. 
 
 $Q1010+0003$ ($z_{em}$ = 1.4007, $z_{abs}$ =1.2651): The Cr II $\lambda$$\lambda$ 2056, 2062, 2066 lines, as well as
 the Zn II $\lambda$$\lambda$ 2026, 2062 lines were detected with S/N of $\sim$ 30. As all the lines were detected, the full
 prescription outlined in $\S$ 2 was followed. The Mg I contribution to the  blended Zn II+Mg I $\lambda$ 2026 line was estimated
 from the SDSS spectrum. The Mg I $\lambda$ 2852 has an equivalent width of W$_{rest}$ = 404 $\mbox{m\AA}$. 
  This line was fit using a 
 single component model, yielding log N$_{\rm Mg I}$ = 12.67$\pm$0.05. This component was held fixed in Mg I $\lambda$ 2026 while 
 the Zn II and Cr II $\lambda$$\lambda$ 2026, 2056, 2062, 2066 lines 
 were fit simultaneously giving log N$_{\rm Zn II}$ = 12.96.

\begin{table*}
{\footnotesize
\caption{Equivalent Widths}
\begin{tabular}{cclcr|cclcr}
\hline
\hline
QSO & $z_{abs}$ & Species & $\lambda_{rest}$ & W$_{rest}$  &   QSO & $z_{abs}$ & Species & $\lambda_{rest}$ & W$_{rest}$  \\
    &           &         &  $\mbox{\AA}$    & m$\mbox{\AA}$ &                &         &  &  $\mbox{\AA}$   & m$\mbox{\AA}$ \\
\hline
0738+313  	&	 0.0912:A   	&	  Ca II       	&	3933	&	  189$\pm$13  	&	  1501+0019 	&	 1.4832:A  	&	 Mg I         	&	2852	&	  952$^{a}$   \\
$z_{em}=0.630$ 	&	          	&	  Ca II       	&	3969	&	  54$\pm$10   	&	 $z_{em}=1.930$    	&	         	&	 Al III       	&	1854	&	  334$\pm$17   \\
          	&	 0.2210:B   	&	  Ca II       	&	3933	&	 63$\pm$10    	&	    	&	         	&	 Al III       	&	1862	&	  216$\pm$18    \\
          	&	          	&	  Ca II       	&	3969	&	   35$\pm$10  	&	       	&	         	&	 Si II        	&	1808	&	     295$\pm$16   \\             
          	&	          	&	  Ti II       	&	3383	&	 $<$14        	&	      	&	         	&	 Cr II        	&	2056	&	  105$\pm$16 \\
0827+243  	&	 0.2590:A  	&	  Ca II       	&	3933	&	 $<$34        	&	        	&	         	&	 Cr II        	&	2066	&	  $<$26    \\
$z_{em}=0.941$ 	&	          	&	  Ca II       	&	3969	&	 $<$34        	&	   	&	         	&	 Zn II+Mg I   	&	2026	&	     231$\pm$21     \\
          	&	          	&	  Ti II       	&	3383	&	 $<$33        	&	     	&	         	&	 Zn II+Cr II  	&	2062	&	   191$\pm$18      \\
          	&	  0.5247:B 	&	 Mg I         	&	2852	&	 602$\pm$29   	&	        	&	 1.8510:B  	&	 C IV         	&	1548	&	  154$\pm$18        \\
          	&	         	&	 Mg II        	&	2796	&	 2416$\pm$27  	&	        	&	         	&	 C IV         	&	1550	&	   116$\pm$18  \\
          	&	         	&	 Mg II        	&	2803	&	  2316$\pm$28 	&	       	&	 1.9260:C  	&	 C IV         	&	1548	&	  254$\pm$14     \\
          	&	         	&	 Ti II        	&	3383	&	 $<$30        	&	       	&		&	 C IV         	&	1550	&	  154$\pm$14    \\ 
1010+0003 	&	 1.2651:A  	&	 Mg I         	&	2852	&	 404$^a$      	&		&		&	 Al II        	&	1670	&	  88$\pm$11 \\
$z_{em}=1.398$ 	&	         	&	 Al III       	&	1854	&	 104$\pm$21   	&	  1712+5559 	&	 1.1590:A  	&	 Cr II        	&	2056	&	  $<$33 \\
          	&	         	&	 Al III       	&	1862	&	 $<$46        	&	  $z_{em}=1.356$      	&	         	&	 Cr II        	&	2066	&	  $<$33    \\
          	&	         	&	 Cr II        	&	2056	&	 91$\pm$19    	&		&		&	 Fe II        	&	2344	&	  415$\pm$30  \\
          	&	         	&	 Cr II        	&	2066	&	 69$\pm$15    	&	      	&		&	 Fe II        	&	2374	&	  240$\pm$27\\
          	&	         	&	 Fe II        	&	2249	&	 144$\pm$26   	&	    	&	         	&	 Fe II        	&	2382	&	  592$\pm$32 \\
          	&	         	&	 Fe II        	&	2260	&	  184$\pm$26  	&	   	&	         	&	 Zn II+Mg I   	&	2026	&	  $<$33 \\
          	&	         	&	 Fe II        	&	2344	&	  564$\pm$28  	&	      	&	         	&	 Zn II+Cr II  	&	2062	&	  $<$33   \\
          	&	         	&	 Fe II        	&	2374	&	  543$\pm$35 	&	   	&	 1.2093:B  	&	 Mg I         	&	2852	&	  367$^{a}$ \\
          	&	         	&	 Fe II        	&	2382	&	   715$\pm$30 	&	     	&	         	&	 Cr II        	&	2056	&	  $<$35 \\
          	&	         	&	 Zn II+Mg I   	&	2026	&	  233$\pm$23  	&	   	&	        	&	 Cr II        	&	2066	&	  $<$35 \\
          	&	         	&	 Zn II+Cr II  	&	2062	&	  181$\pm$22  	&	   	&	          	&	 Fe II        	&	2344	&	  973$\pm$33 \\
1107+0003 	&	 0.5252:A  	&	 Mn II        	&	2576	&	 $<$30        	&	    	&	        	&	 Fe II        	&	2374	&	  723$\pm$31  \\
$z_{em}=1.741$  	&		&	 Mn II        	&	2594	&	 $<$34        	&	    	&	         	&	 Fe II        	&	2382	&	  1168$\pm$35    \\
          	&	         	&	 Mn II        	&	2606	&	 $<$20        	&		&		&	 Zn II+Mg I   	&	2026	&	  $<$35\\
          	&	         	&	 Fe II        	&	2344	&	 139$\pm$36   	&		&		&	 Zn II+Cr II  	&	2062	&	  $<$35 \\
          	&	         	&	 Fe II        	&	2374	&	 $<$47        	&	  1715+4606              	&	 GAL$^{b}$     	&	 Ca II   	&	3933	&	  1556$\pm$46$^{c}$ \\
          	&	         	&	 Fe II        	&	2382	&	  269$\pm$36  	&	  $z_{em}=0.989$     	&		&	 Ca II    	&	3969	&	   205$\pm$36  \\
          	&	         	&	 Fe II        	&	2586	&	  185$\pm$25  	&	  	&	 0.6544:A  	&	 Cr II        	&	2056	&	  $<$165  \\
          	&	         	&	 Fe II        	&	2600	&	   342$\pm$27 	&		&		&	 Cr II        	&	2066	&	  $<$165    \\
          	&	 0.9547:B  	&	 Ti II        	&	1910	&	 $<$42        	&		&		&	 Fe II        	&	2344	&	  1264$\pm$42\\ 
          	&	         	&	 Cr II        	&	2056	&	 $<$27        	&	   	&	         	&	 Fe II        	&	2374	&	  851$\pm$40\\
          	&	         	&	 Cr II        	&	2066	&	 $<$27        	&	  	&		&	 Fe II        	&	2382	&	    1556$\pm$46$^{c}$  \\
          	&	         	&	 Zn II+Mg I   	&	2026	&	 $<$27        	&	   	&	         	&	 Zn II+Mg I   	&	2026	&	  $<$165 \\
          	&	         	&	 Zn II+Cr II  	&	2062	&	 $<$27        	&	   	&	         	&	 Zn II+Cr II  	&	2062	&	  $<$165  \\
          	&	 1.711:C   	&	 Si IV        	&	1393	&	  618$\pm$36  	&	 1733+5533  	&   GAL$^{b}$ 		&	 Ca II         	&	3933	&	 482$\pm$36\\
          	&	         	&	 Si IV        	&	1402	&	  447$\pm$40  	&	  $z_{em}=1.074$  	&	 	&	 Ca II         	&	3969	&	  257$\pm$33\\
1137+3907 	&	 GAL$^{b}$     	&	Ca II    	&	3933	&	336$\pm$58	&	  	&	   	&	 Mg I         	&	2852	&	  362$^{a}$ \\
$z_{em}=1.023$  	&	        	&	Ca II	&	3969	&	241$\pm$51	&	   	&	 0.9984:A  	&	 Al III       	&	1854	&	 179$\pm$31\\
	&	 0.7190:A  	&	 Cr II        	&	2056	&	 $<$74        	&		&		&	 Al III       	&	1862	&	 100$\pm$27   \\
	&	         	&	 Cr II        	&	2066	&	 $<$74        	&		&		&	 Si II        	&	1808	&	  178$\pm$30    \\
	&		&	 Fe II        	&	2249	&	  150$\pm$31  	&		&		&	 Ti II        	&	1910	&	  $<$30    \\
          	&	         	&	 Fe II        	&	2260	&	  291$\pm$34  	&		&		&	 Cr II        	&	2056	&	 $<$35    \\
          	&	         	&	 Fe II        	&	2344	&	 1768$\pm$67  	&		&		&	 Cr II        	&	2066	&	  $<$29\\
          	&	         	&	 Fe II        	&	2374	&	 1300$\pm$37  	&	   	&	         	&	 Zn II+Mg I    	&	2026	&	  $<$29\\
          	&	         	&	 Fe II        	&	2382	&	  2009$\pm$70 	&	   	&	         	&	 Zn II+Cr II    	&	2062	&	  $<$29\\
	&		&	 Zn II+Mg I   	&	2026	&	 447$\pm$56   	&	   	&	 1.1496:B    	&	 C IV          	&	1548	&	  453$\pm$55 \\
	&		&	 Zn II+Cr II  	&	2062	&	 266$\pm$70   	&	  	&	         	&	 C IV          	&	1550	&	  341$\pm$51  \\
1225+0035 	&	 0.7731:A  	&	 Mg I         	&	2852	&	 929$^{a}$    	&	  	&	          	&	 Al II         	&	1670	&	  $<$40\\
$z_{em}=1.226$ 	&	           	&	 Cr II        	&	2056	&	   259$\pm$81 	&	  	&	         	&	 Al III        	&	1854	&	  $<$30\\
          	&	         	&	 Cr II        	&	2066	&	   177$\pm$76  	&	  	&	         	&	 Al III        	&	1862	&	  $<$30    \\
          	&	         	&	 Fe II        	&	2249	&	  339$\pm$73  	&	  	&	         	&	 Si II         	&	1808	&	  $<$25\\
          	&	         	&	 Fe II        	&	2260	&	    372$\pm$80 	&	$\cdots$ 	&	$\cdots$ 	&	$\cdots$ 	&	$\cdots$ 	&	$\cdots$ \\
	       &		&	 Zn II+Mg I   	&	2026	&	  $<$122         	&	$\cdots$ 	&	$\cdots$ 	&	$\cdots$ 	&	$\cdots$ 	      &	$\cdots$       \\
	       &		&	 Zn II+Cr II  	&	2062	&	   181$\pm$91    	&	$\cdots$ 	&	$\cdots$ 	&	$\cdots$ 	&	$\cdots$ 	      &	$\cdots$         \\
\hline
\end{tabular}
\vspace{0.2cm}
\begin{minipage}{140mm}
{\bf $^a$:}  From SDSS spectra.
{\bf $^b$:}  Entries with GAL identifier are from lines originating in the Milky Way.
{\bf $^c$:}  The Galactic Ca II $\lambda$ 3933 line and Fe II $\lambda$ 2383 line from the DLA at 
z$_{abs}$ = 0.6544 are blended in the spectra.
\end{minipage}
}
\end{table*}

\begin{figure*}
\includegraphics[width=7.0in,height=8.5in,angle=180]{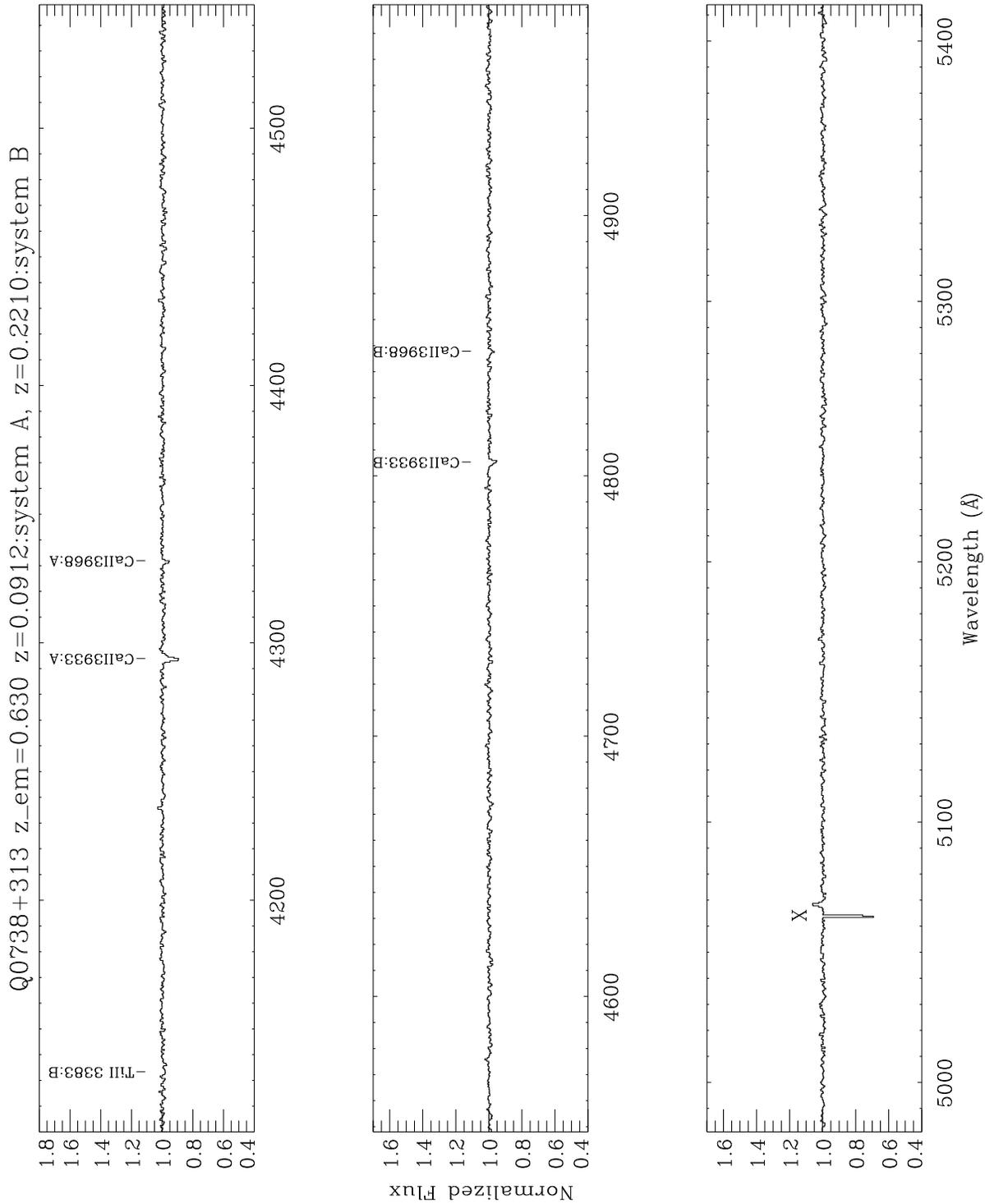}
\caption{Combined and normalized spectrum of Q0738+313. The expected positions of several lines of interest are 
marked above the continuum with their rest wavelength and species. The letters after the wavelengths refer to 
the different systems indicated at the top of the figure. An ``X'' signifies a defect in the CCD.}
\end{figure*}
\clearpage

\begin{figure*}
\includegraphics[width=7.0in,height=8.5in,angle=180]{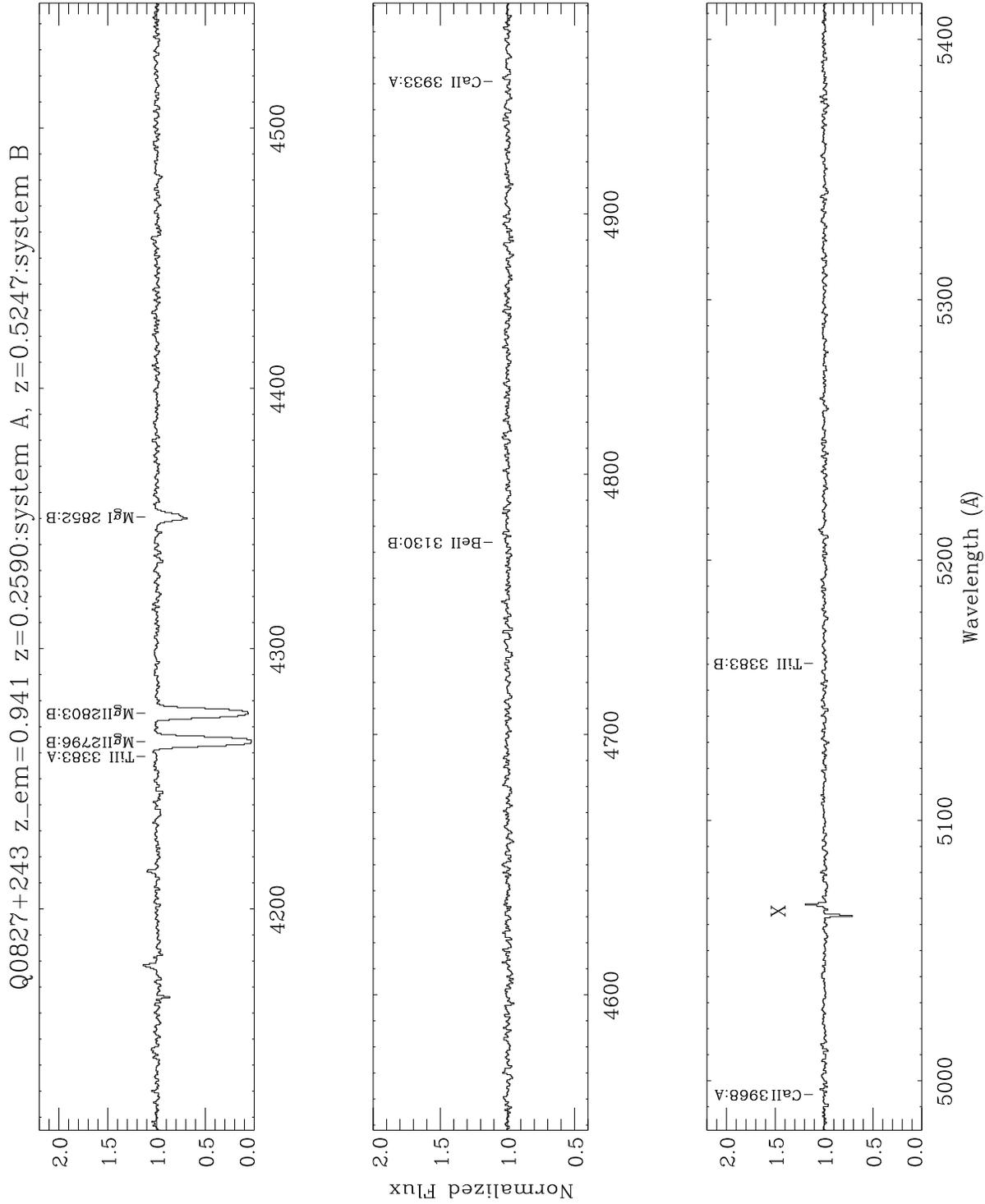}
\caption{Same as Figure 1, but for Q0827+243.}
\end{figure*}
\clearpage

\begin{figure*}
\includegraphics[width=7.0in,height=8.5in,angle=180]{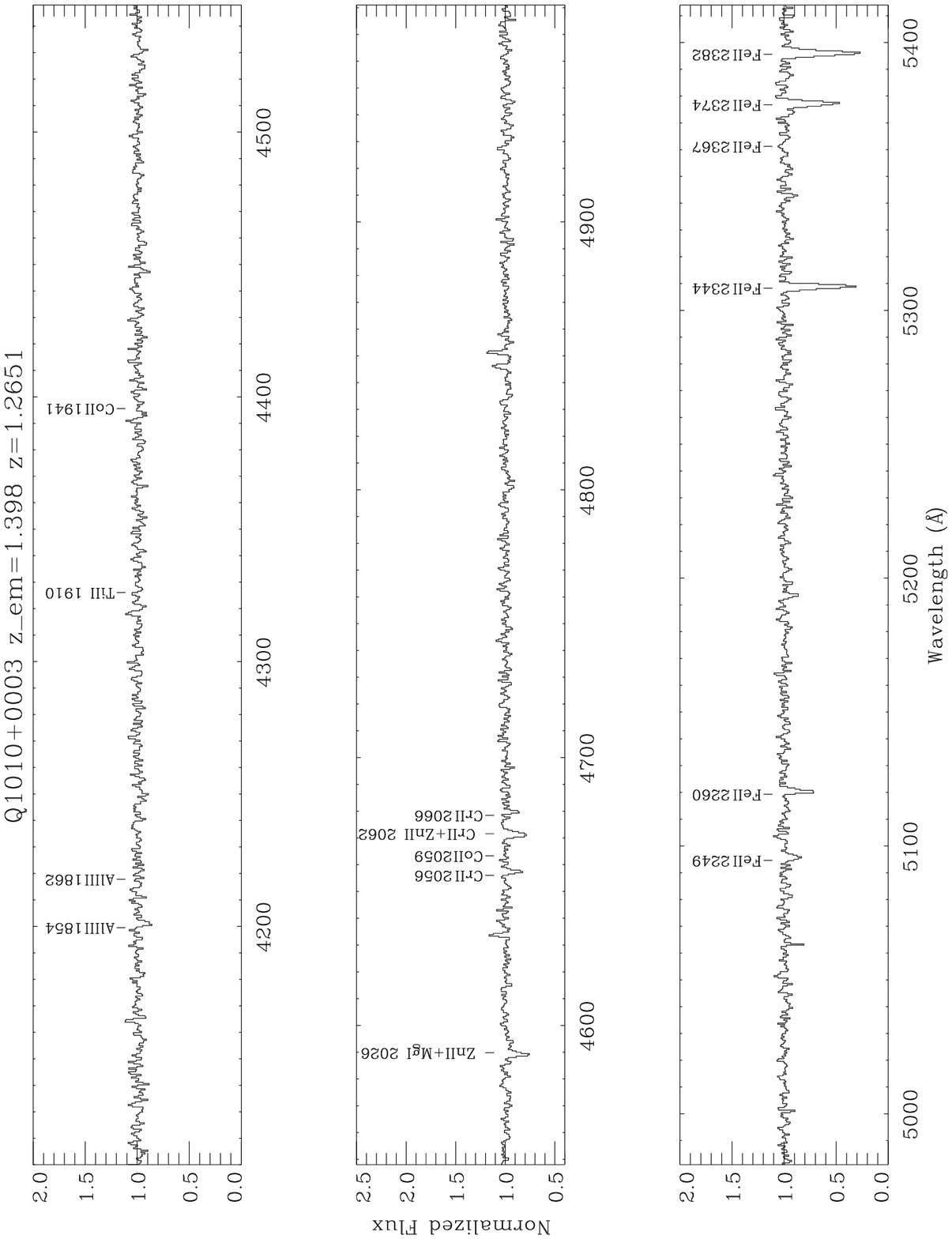}
\caption{Same as Figure 1, but for Q1010+0003.}
\end{figure*}
\clearpage

\begin{figure*}
\includegraphics[width=7.0in,height=8.5in,angle=180]{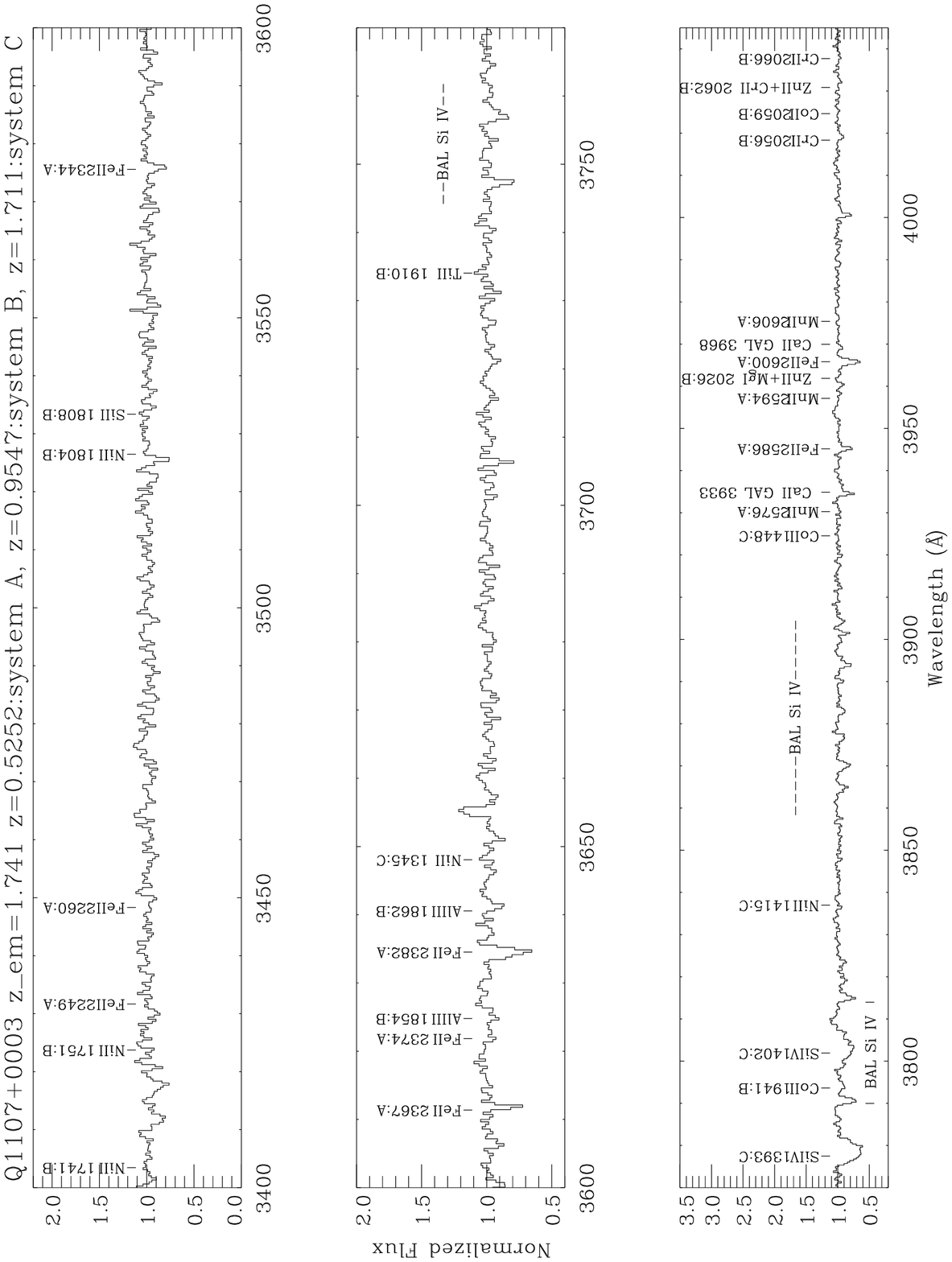}
\caption{Same as Figure 1, but for Q1107+0003.}
\end{figure*}
\clearpage

\begin{figure*}
\includegraphics[width=7.0in,height=8.5in,angle=180]{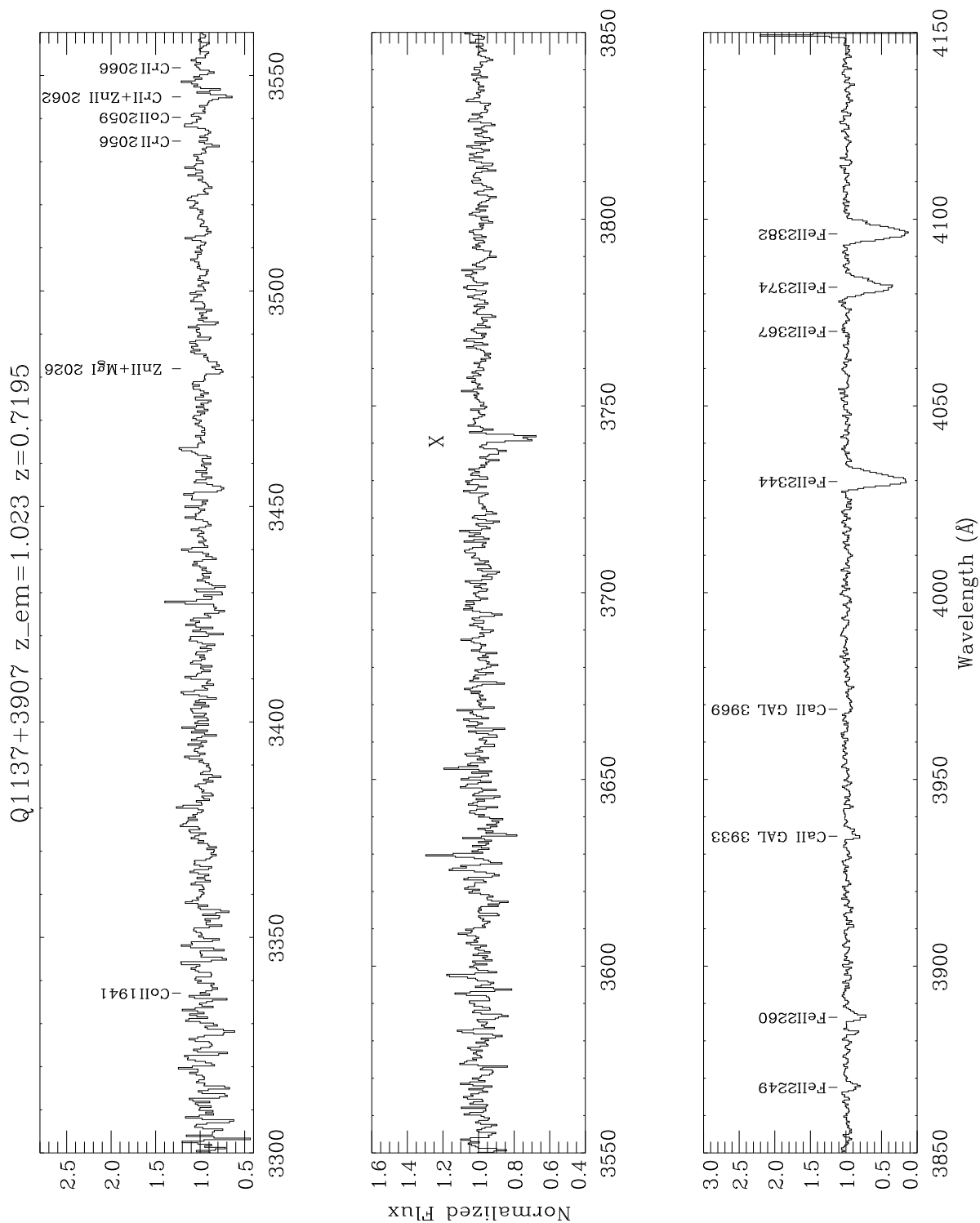}
\caption{Same as Figure 1, but for Q1137+3907.}
\end{figure*}
\clearpage

\begin{figure*}
\includegraphics[width=7.0in,height=8.5in,angle=180]{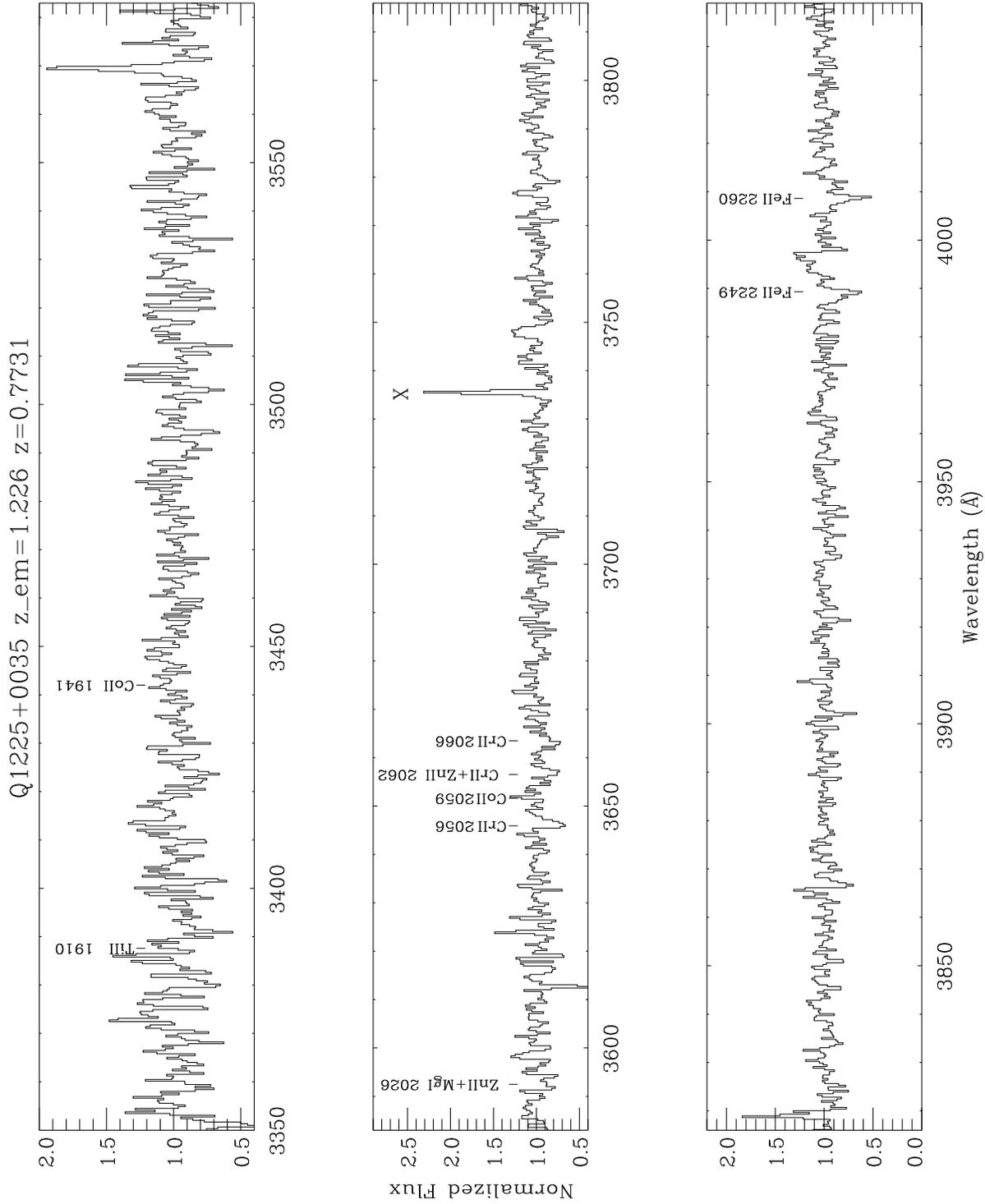}
\caption{Same as Figure 1, but for Q1225+0035.}
\end{figure*}
\clearpage

\begin{figure*}
\includegraphics[width=7.0in,height=8.5in,angle=180]{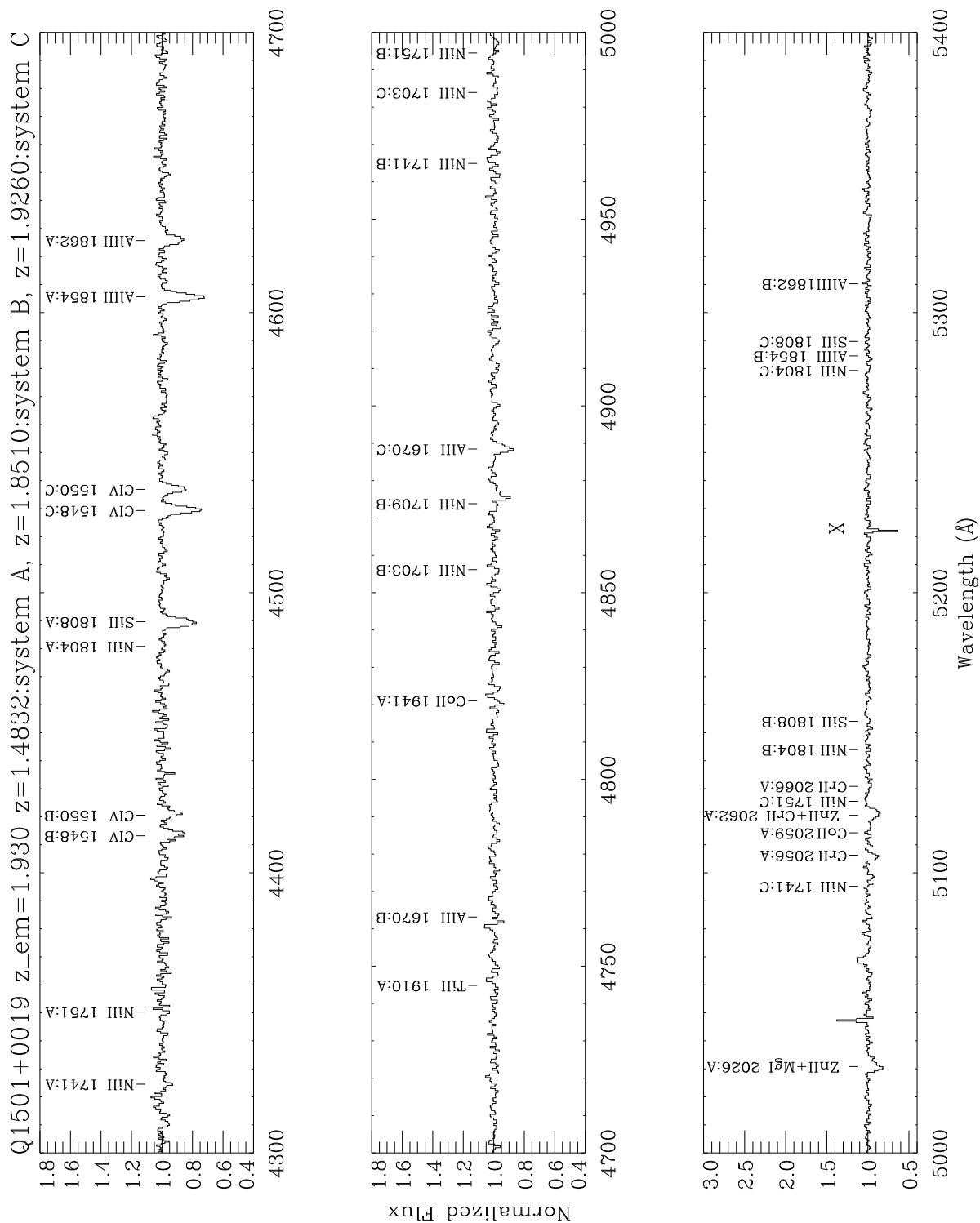}
\caption{Same as Figure 1, but for Q1501+0019.}
\end{figure*}
\clearpage

\begin{figure*}
\includegraphics[width=7.0in,height=8.5in,angle=180]{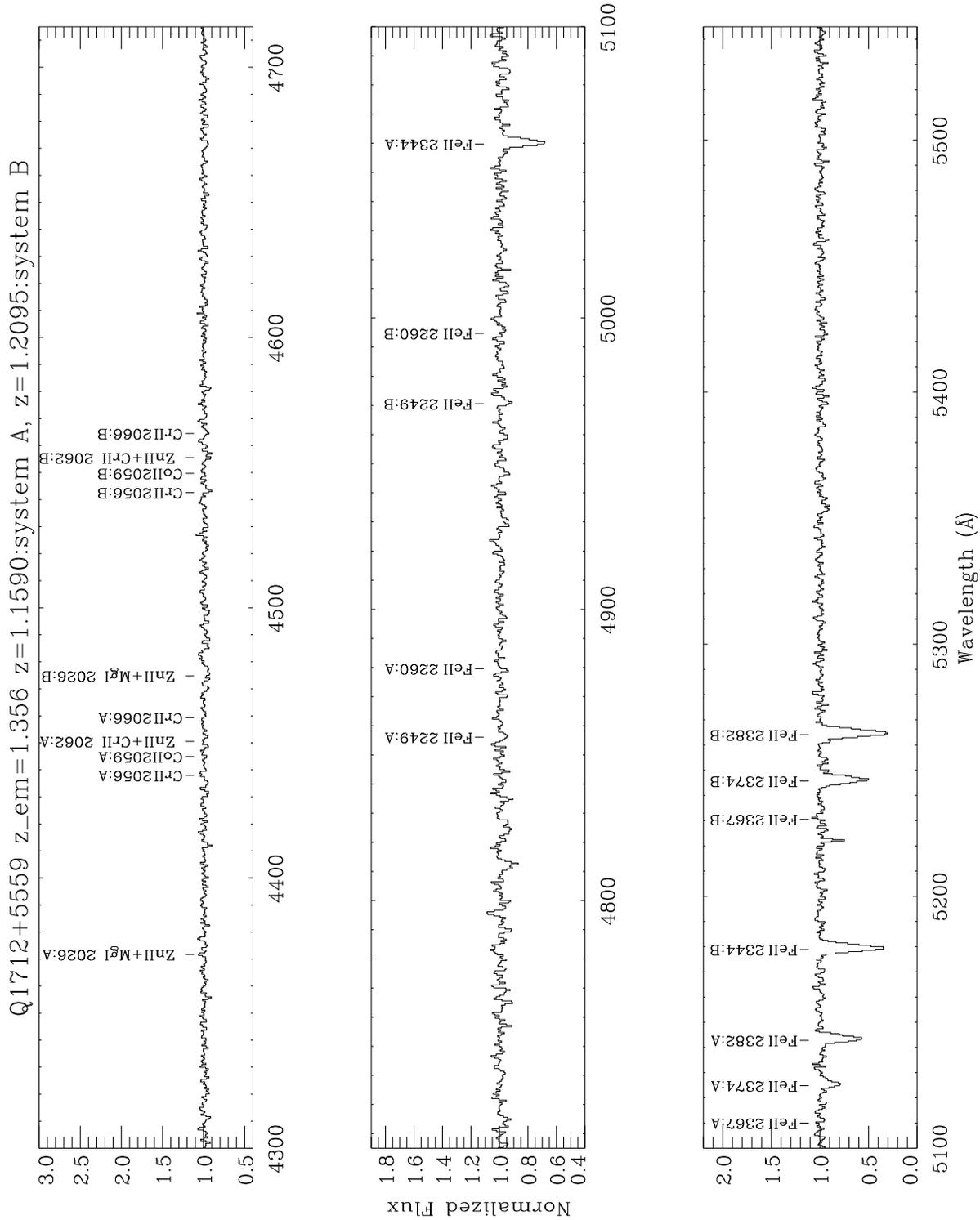}
\caption{Same as Figure 1, but for Q1712+5559.}
\end{figure*}
\clearpage

\begin{figure*}
\includegraphics[width=7.0in,height=8.5in,angle=180]{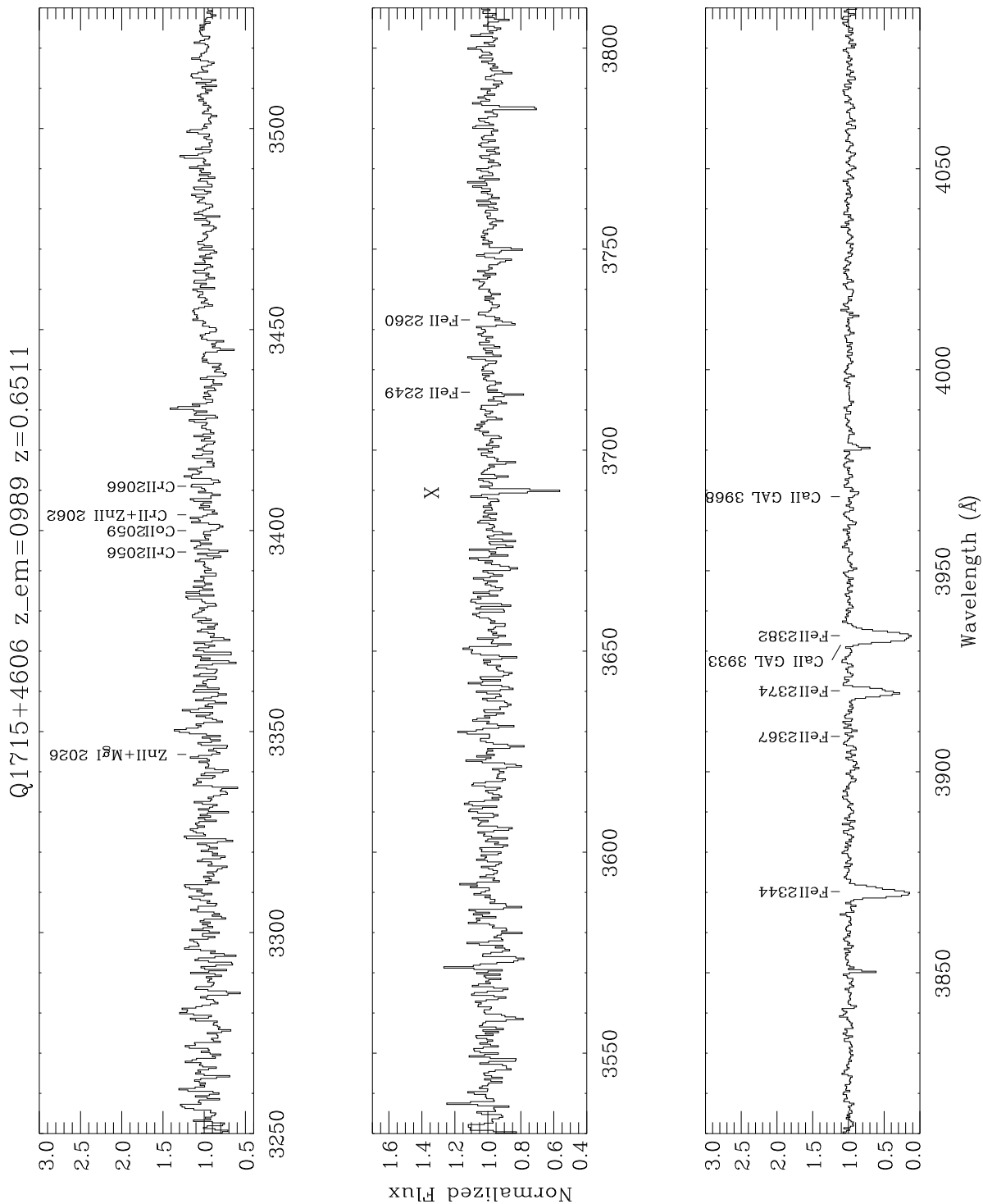}
\caption{Same as Figure 1, but for Q1715+4606.}
\end{figure*}
\clearpage

\begin{figure*}
\includegraphics[width=7.0in,height=8.5in,angle=180]{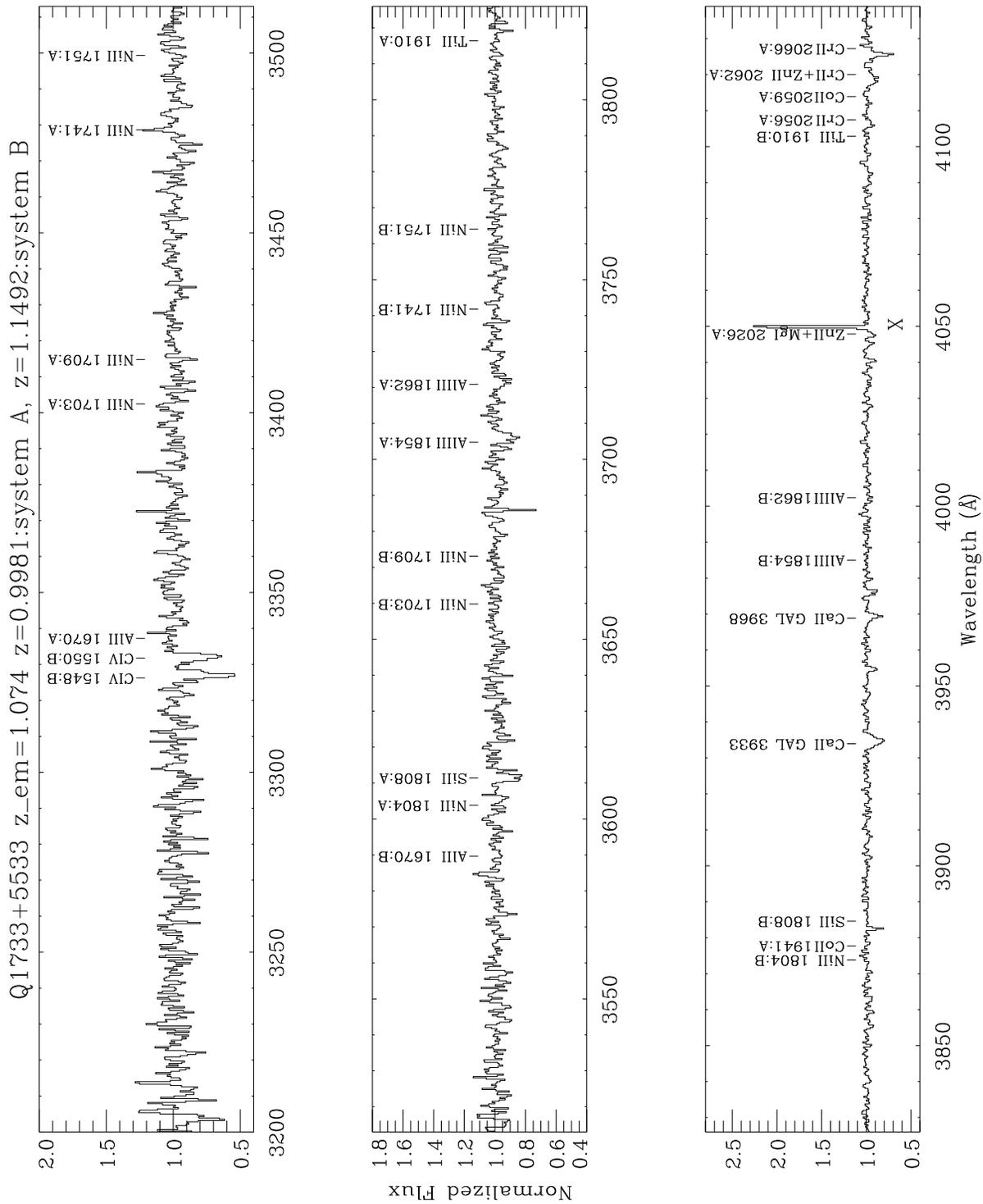}
\caption{Same as Figure 1, but for Q1733+5533.}
\end{figure*}
\clearpage

------------

To judge the effect of potential saturation of Mg I $\lambda$ 2852 we estimated 
 the maximum N$_{\rm Mg I}$ and hence the minimum N$_{\rm Zn II}$ as described in $\S$ 3. 
 This gave log N$_{\rm Zn II}$ $>$ 12.95, affirming the assumption that the Mg I 
 contribution to the blended $\lambda$ 2026 line is small. Fe II $\lambda$$\lambda$ 2249, 2260, 2344, 2374, 2382 lines were also detected.

$Q1107+0003$ ($z_{em}$ = 1.7408, $z_{abs}$ = 0.5252 for system A, $z_{abs}$ = 0.9547 for system B and $z_{abs}$ = 1.711 for system C):
 System B is of primary interest to this paper, and contains a confirmed sub-DLA from HST spectra with log N$_{\rm H I}$ = 20.26.
 This QSO also has a broad absorption line system (BAL) at $z$ $\approx$ 1.711. Si IV $\lambda$$\lambda$ 1393, 1403 lines from the BAL
 were detected. No Zn II or Cr II lines were detected from the sub-DLA at $z$ = 0.9547, but upper limits were placed on the column
 densities. There is also a system at $z$ = 0.5252 with weak Fe II lines detected in SDSS spectra. Lines of Fe II $\lambda$$\lambda$ 2344,
 2382, 2586, 2600 were detected from this system. The weaker Fe II $\lambda$ 2374 line however was not detected.  The Fe II $\lambda$
 $\lambda$ 2344, 2382 lines were fit simultaneously, as were the $\lambda$$\lambda$ 2586, 2600 lines with the same $b_{eff}$ and $v$. 
 Mn II $\lambda$$\lambda$ 2576, 2594, 2606 fell within the observable region for system B, but none were
 detected.

$Q1137+3907$ ($z_{em}$ = 1.023 and $z_{abs}$ = 0.7193): This is a BAL system. The Zn II + Mg I $\lambda$ 2026 blend was detected at $\sim$ 3$\sigma$ level in
 this system. The S/N in the region was $\sim$ 15, leading to larger errors in the column density estimates.  There are no data available
 on Mg I for this system, so it could not be included in the fit of the blended Zn II+Mg I $\lambda$ 2026. We are however confident of
 the Zn II column density quoted, because based on our experience so far, the contribution from Mg I $\lambda$ 2026 is relatively small
 for most systems.  For example, the $z = 1.2651$ DLA toward Q1010+0003 showed differences in the Zn II column density of only $\sim$ 0.05
 dex with and without the contribution from Mg I $\lambda$ 2026.The Zn II $\lambda$ 2026 line was fit using the $b_{eff}$ value from the fit of the Fe II $\lambda$$\lambda$ 2249, 2260
 lines. The Zn II component was held fixed in the blended $\lambda$ 2062 line to determine N$_{\rm Cr II}$.
 Strong Fe II $\lambda$$\lambda$ 2249, 2260, 2344, 2374, 2382 were also
 detected. The feature at $\sim$ 3740 $\mbox{\AA}$ is caused by a defect in the CCD.

$Q1225+0035$ ($z_{em}$ = 1.226 and $z_{abs}$ = 0.7730): The spectra of this object were taken near daybreak, which resulted in
 additional noise due to increased background. The Cr II
 $\lambda$$\lambda$ 2056, 2062 lines was detected at $\sim$ 3$\sigma$, allowing for moderately accurate determination of the column
 density. The Zn II+Mg I $\lambda$ 2026 blend lies in a noisy region and was not detected, and the Zn II upper limit had to be 
 estimated from S/N. We also detected Fe II $\lambda$ 2249, 2260 lines which were fit simultaneously with the same $b_{eff}$ and $v$. 
 The feature at $\sim$ 3735 $\mbox{\AA}$ is caused by a defect in the CCD.
 
$Q1501+0019$ ($z_{em}$ = 1.930, $z_{abs}$ = 1.4832 for system A, $z_{abs}$ = 1.8510 for system B, and $z_{abs}$ = 1.9260 for system C):
 System A is the system of primary interest in this paper, and has a confirmed DLA from HST spectra. We detected only the C IV            
 $\lambda$$\lambda$ 1548, 1550 lines in system B.  The Zn II+Mg I $\lambda$ 2026 blend, along with the Cr II+Zn II $\lambda$ 2062 and Cr
 II $\lambda$ 2056 lines were all detected in
 system A. Because the Cr II $\lambda$ 2066 line was not detected, the Cr II column density was determined from Cr II $\lambda$ 2056
 solely. The Mg I contribution to the blended Zn II+Mg I  $\lambda$ 2026 line was determined from SDSS spectra, which yields a W$_{rest}$
 for the Mg I  $\lambda$ 2852 line of 952 $\mbox{m\AA}$. This line was fit using a one component model, giving 
 log N$_{\rm Mg I}$ = 12.93$\pm$0.05. This component was held fixed in the MMT spectrum of the $\lambda$ 2026 line
 while the Zn II and Cr II $\lambda$$\lambda$ 2056, 2062, and 2066 lines were fit simultaneously. 
 To judge the effects of saturation of Mg I $\lambda$ 2852, we again estimated the maximum Mg I contribution to the blended
 $\lambda$ 2026 line as described in $\S$ 3. This gave N$_{\rm Zn II}$ $>$ 12.89, only 0.04 dex less than the ``best fit'' model.
 Al III $\lambda$$\lambda$ 1854, 1862 lines were also detected and
 fit together while simultaneously varying N, $b_{eff}$, and $v$. The feature at $\sim$ 5220 $\mbox{\AA}$ is caused by a defect in the CCD.

$Q1712+5559$ ($z_{em}$ = 1.930, $z_{abs}$ = 1.2093 for system A and $z_{abs}$ = 1.1584 system B): System A is the system of primary
 interest  for this paper, and contains a confirmed DLA from HST spectra with log N$_{\rm H I}$ = 20.72. Both systems contain strong Fe
 II  $\lambda$$\lambda$ 2344, 2374, 2382 lines. No other lines from system B appear to be present. No Zn II or Cr II lines were
 detected in either system, with S/N $\sim$ 30 in the region. The feature at $\sim$ 5220 $\mbox{\AA}$ is caused by a defect in the CCD.

$Q1715+4606$ ($z_{em}$ = 0.989 and $z_{abs}$ = 0.6511)  We have determined the H I column density for this object from archival $HST$
 data. By fitting the line profile, we have determined a column density of log N$_{\rm H I}$=20.44. Strong Fe II $\lambda$$\lambda$
 2344, 2374, 2382 lines were detected. No lines of Zn II or Cr II were seen, although the S/N in the region was only $\sim$ 10 due to poor
 weather during the observations. A weak Galactic Ca II $\lambda$ 3698 feature was seen, but the stronger Ca II $\lambda$ 3933 line was
 blended with the Fe II $\lambda$ 2382 line from the $z_{abs}$=0.6511 system. The feature located at $\sim$ 3690 $\mbox{\AA}$ is an artifact of the CCD.

$Q1733+5533$ ($z_{em}$ = 1.074 and $z_{abs}$ = 0.9984 for system A and $z_{abs}$ = 1.1496 for system B) System A is of primary interest
 for this study, as it has a confirmed DLA with log N$_{\rm H I}$ = 20.70. System B appears to be a C IV system with no other
 absorption features at this redshift detected in SDSS spectra. No Zn II or Cr II lines were detected in system A, but the relatively
 high S/N ($\sim$33) in the region allows us to place tight upper limits on the abundances. Al III $\lambda$$\lambda$ 1854,1862 lines
 were weakly detected and were fit together varying N, $b_{eff}$, and $v$ simultaneously. We also detected strong Galactic Ca II $\lambda$
$\lambda$ 3933, 3968 lines.

\begin{table*}
{\footnotesize
\begin{center}
\caption{Column Densities }
\begin{tabular}{ccccl|ccccl}
\hline
\hline
QSO & $z_{abs}$ & Species & log N     & $b_{eff}$  & QSO    &  $z_{abs}$ & Species & log N & $b_{eff}$       \\
    &           &         & cm$^{-2}$ &  km s$^{-1}$  &    &             &        &  cm$^{-2}$ & km s$^{-1}$ \\
\hline 
0738+313 	&	0.0912	&	 Ca II 	&	 12.32$\pm$0.02  	&	51.2	&	1225+0035 	&		0.7731	&	 Mg I  	&	 12.85$\pm$0.06$^{c}$    &	52.4 \\
          	&	          	&	 Ti II 	&	 12.53$\pm$0.10$^{a}$ 	&	 $\cdots$             	&	&         	&	 Cr II 	&	 13.99$\pm$0.11 	&	74.7 \\
          	&	          	&	 Cr II 	&	 13.28$\pm$0.22$^{b}$ 	&	 $\cdots$  	&		&	   	&	 Fe II 	&	 15.69$\pm$0.03 	&	  74.7  \\
          	&	          	&	 Zn II 	&	 $<$12.66$^{a}$   	&	 $\cdots$     	&	  	&	       	&	 Zn II 	&	 $<$13.01    		&	 $\cdots$     \\ 
          	&	0.2210	&	 Mg I  	&	 12.00$\pm$0.10$^{a}$ 	&	 $\cdots$    	&	 1501+0019 	&	1.4832	&	 Mg I  	&	  12.92$\pm$0.05$^{c}$  &	 96.6     \\
          	&	          	&	 Mg II 	&	 $>$13.30$\pm$0.03$^{a}$ 	&	 $\cdots$ 	&	  	&	&	 Al III	&	 13.35$\pm$0.01   	&	  87.1     \\
          	&	          	&	 Ca II 	&	 11.91$\pm$0.03 	&	34.2	&	  	&	         	&	 Si II 	&	 15.71$\pm$.02  	&	 86.5     \\
          	&	          	&	 Ti II 	&	 $<$11.48  	&	 $\cdots$         	&	  	&	        &	 Cr II 	&	 13.40$\pm$0.09  	&	 86.5  \\
          	&	          	&	 Cr II 	&	  13.11$\pm$0.24$^{b}$ 	&	 $\cdots$   	&	              &	   	&	 Zn II 	&	 12.93$\pm$0.06  	&	     86.5                \\
          	&	          	&	 Zn II 	&	  $<$12.83$^{b}$  	&	 $\cdots$      	&	     	&	1.8510	&	 C IV  	&	 13.64$\pm$0.04  	&	  74.1      \\
0827+243 	&	0.2590	&	 Ca II 	&	  $<$11.60     	&	  $\cdots$       	&	      	&	1.9260		&	 C IV  	&	  13.87$\pm$0.03 	&	  88.8     \\     
          	&	          	&	 Ti II 	&	  $<$11.96     	&	 $\cdots$        	&	        &        	&	 Al II 	&	  12.33$\pm$0.04 	&	  53.4     \\ 
          	&	0.5247	&	 Mg I  	&	 12.69$\pm$0.02 	&	80.3	&	 1712+5559 	&	1.1584		&	 Cr II 	&	 $<$12.93    		&	  $\cdots$  \\
          	&	          	&	 Mg II 	&	 $>$14.59   	&	80.3	&	     	&	           		&	 Fe II 	&	 13.98$\pm$0.03 	&	 74.9 \\
          	&	          	&	 Ti II 	&	 $<$11.76    	&	 $\cdots$     	&	        	&	   	&	 Zn II 	&	 $<$12.25       	&	  $\cdots$           \\    
          	&	          	&	 Cr II 	&	 $<$13.42$^{b}$ 	&	 $\cdots$     	&	    	&	1.2093	&	 Mg I  	&	 12.44$^{c}$     	&	  76.7      \\
	&		&	Fe II	& 	14.59$\pm$0.02$^{a}$	&	 $\cdots$     	&	  	&	         		&	 Cr II 	&	 $<$12.86       	&	  $\cdots$      \\
          	&	          	&	 Zn II 	&	 $<$12.80$^{b}$ 	&	 $\cdots$  	&	  	&	        &	 Fe II 	&	 $>$14.54 		&	  76.7       \\
1010+0003 	&	1.2651	&	 Mg I  	&	  12.67$\pm$0.05$^{c}$  	&	184.3	&	  	&	         	&	 Zn II 	&	 $<$12.19        	&	  $\cdots$            \\
          	&	          	&	 Al III &	 12.67$\pm$0.10 	&	10.8	&	    1715+4606  	&GAL$^{d}$ 	&	 Ca II 	&	 12.66$\pm$0.10	 	&	 35.5 \\
          	&	          	&	 Cr II 	&	 13.54$\pm$0.07 	&	22.5	&	        	&	0.6511	&	 Cr II 	&	 $<$13.54        	&	$\cdots$      \\
          	&	          	&	 Fe II 	&	 15.26$\pm$0.05 	&	22.5	&	  	&	         	&	 Fe II 	&	 14.94$\pm$0.03  	&	  57.4        \\
          	&	          	&	 Zn II 	&	 12.96$\pm$0.06 	&	22.5	&	  	&	         	&	 Zn II 	&	 $<$12.87        	&	$\cdots$        \\  
1107+0003 	&	0.5252	&	 Mn II 	&	 $<$12.16    	&	$\cdots$	&	 1733+5533  	&	 GAL$^{d}$    	&	 Ca II 	&	  12.79$\pm$0.03 	&	 88.2 \\
          	&	          	&	 Fe II 	&	 13.54$\pm$0.08 	&	37	&	    	&	0.9984		&	 Mg I  	&	   12.44$^{b}$ 		&       $\cdots$ 	   \\  
          	&	0.9547	&	 Ti II 	&	 $<$13.01  	&	 $\cdots$       	&	    	&	        	&	 Al III	&	 13.13$\pm$0.05 	&	133.7\\
          	&	          	&	 Cr II 	&	 $<$12.76  	&	 $\cdots$       	&	  	&	       	&	 Si II 	&	 15.48$\pm$0.06 	&	118.9\\
          	&	          	&	 Zn II 	&	 $<$12.08  	&	  $\cdots$     	&	  	&	        	&	 Cr II 	&	 $<$12.79      		&	$\cdots$   \\
          	&	1.7110	&	 Si IV 	&	 13.99$\pm$0.03 	&	198.3	&	   	&	        		&	 Zn II 	&	 $<$12.11     		&	$\cdots$    \\
1137+3907 	&	 GAL$^{d}$ 	&	Ca II	&	12.63$\pm$0.05	&	87.8	&	  	&	1.1496			&	 C IV  	&	  14.19$\pm$0.04 	&	 88.8 \\
          	&	0.7193	&	 Cr II 	&	 13.71$\pm$0.20 	&	122.9	&	  	&	          		&	 Al II 	&	  $<$11.77    		&	$\cdots$            \\
          	&	          	&	 Fe II 	&	 15.45$\pm$0.05 	&	122.9	&	       	&	   		&	 Si II 	&	  $<$14.6    	 	&	$\cdots$        \\
	&		&	 Zn II 	&	 13.43$\pm$0.05 	&	122.9	&	  $\cdots$     	&	  $\cdots$     		& $\cdots$     	&	  $\cdots$   	  	&	$\cdots$ \\

\hline
\end{tabular}
\end{center}
\vspace{0.2cm}
\begin{minipage}{140mm}
{\bf $^a$:}  Khare et al. (2004)
{\bf $^b$:}  Kulkarni et al. (2005)
{\bf $^c$:}  From SDSS spectra.
{\bf $^d$:}  Entries with GAL identifier are from lines originating in the Milky Way.
\end{minipage}
}
\end{table*}


\begin{table*}
\begin{center}
\caption{Relative Abundances
\label{t:JoO}}
\begin{tabular}{ccccccc}
\hline
\hline
QSO & $z_{abs}$ & [Zn/H]  & [Cr/Zn] & [Fe/Zn] & [Ti/Zn] & [Si/Zn] \\
\hline
Q0738+313  & 0.0912 & $<$-1.14$^{a}$ & $>$-0.21$^{a}$ & $>$-0.48$^{a}$ & $\cdots$   & $\cdots$        \\
 $\cdots$  & 0.2213 & $<$-0.7$^{a}$  &  $<$0.7$^{a}$  & $\cdots$ & $\cdots$         & $\cdots$        \\
Q0827+243  & 0.5247 & $<$-0.04$^{a}$ & $<$-0.3$^{a}$  & $\cdots$ & $\cdots$         & $\cdots$       \\
Q1010+0003 & 1.2651 & -1.19          &    -0.46       & -0.54    & $<$-1.01         & $\cdots$      \\
Q1107+0003 & 0.9547 & $<$-0.72       &  $\cdots$      & $\cdots$ & $\cdots$         & $\cdots$         \\
Q1137+3907 & 0.7190 & -0.30          &    -0.67       &  -0.82   & $\cdots$         & $\cdots$         \\
Q1225+0035 & 0.7731 & $<$-1.15       &    $>$0.09      & $>$0.07  & $\cdots$         & $\cdots$          \\
Q1501+0019 & 1.4832 & -0.54         &    -0.56       & $\cdots$ & $<$-1.24 &  -0.28        \\
Q1712+5559 & 1.2093 & $<$-1.07       &    $\cdots$    & $>$-0.51 &  $\cdots$        & $\cdots$          \\
Q1715+4606 & 0.6511 & $<$-0.61       &    $\cdots$    & $>$-0.78 & $\cdots$         &  $\cdots$         \\
Q1733+5533 & 0.9981 & $<$-1.13       &    $\cdots$    & $\cdots$ & $\cdots$        &  $>$0.37         \\              
\hline
\end{tabular}
\end{center}
\vspace{0.2cm}
\begin{minipage}{140mm}
{\bf $^a$:} Abundance measurements from Kulkarni et al. (2005). 
\end{minipage}
\end{table*}

\section{Results and Discussion}
\subsection{Relative Abundances}

The observed relative abundances of the elements are a combination of both the nucleosynthetic processes and of dust depletion. If
 dust is significantly present in DLAs, then refractory elements such as Cr, Mn, Fe, Co, and Ni should show substantial depletions. As 
 already discussed in $\S$ 1, Zn is a relatively undepleted element due to its low condensation temperature. Ratios such as [Cr/Zn]
 and [Fe/Zn] are therefore a measure of dust depletion. Table 5 lists abundances of Cr, Fe, and Ti relative to Zn for the systems in our
 sample. Also given is the metallicity [Zn/H]. Although it is customary to estimate metallicity  based on measurements of 
 N$_{\rm Zn II}$, it has been shown that [Zn/H] is not much affected by ionization \citep{Vlad01}.

   No Ti II was found in any of the systems sampled. We determined [Ti/Zn] upper limits of -1.01 and -1.24 for the DLAs toward Q1010+0003
and Q1501+0019 respectively. The measured [Fe/Zn] in Q1010+0003 was also sub-solar (-0.69) also
 suggesting dust depletion. No Fe lines were sampled in the spectra of Q1501+0019, so [Fe/Zn] could not be measured. 
  
 Si is an $\alpha$-capture element, and its abundance relative to Fe group elements can provide clues into the chemical evolution
 of DLAs. Disk stars in the Milky Way show a systematic decrease in [$\alpha$/Fe] with increasing [Fe/H], which indicates an increase with
 time of Fe contributed to the ISM from type Ia supernovae, relative to type II supernovae \citep{Edv93}. 
 Prochaska $\&$ Wolfe (1999) observed that the DLAs in their sample showed no such trend, and that most of the objects were
 $\alpha$-enhanced. In DLAs, Zn is a better
 indicator of Fe group abundances than Fe because of the large depletion of Fe on dust grains. We detected Si II $\lambda$ 1808 in 
 Q1501+0019 and Q1733+5533. [Si/H] = -0.68 in Q1501+0019, and [Si/H] = -0.76 in Q1733+5533. We unfortunately did not cover any Fe II
 lines in the region. However, [Si/Zn] = -0.28 in Q1501+0019, and [Si/Zn] $>$ 0.37 in Q1733+5533, implying significant   
 $\alpha$-enhancement in the latter system.

Figure 11 shows [Cr/Zn] vs. [Zn/H] for our data as well as those from the literature. Only detections, not limits have been plotted. 
 There is intrinsic scatter in the data, which is to be expected, but there appears to be a general trend suggesting that
 the most metal rich DLAs also show the most depletion. If this is true, then one would expect the most metal rich absorbers to be
 under-sampled because they would also most likely be the most reddened. A Spearman rank-order correlation test on the 48 data points
 plotted in Fig. 11 gives the Spearman rank-order coefficient R$_{\rm S}$=-0.623, with a probability of obtaining such a value by chance
 of  P(R$_{\rm S}$)=1.95$\times$10$^{-5}$.

 Figure 12 shows [Fe/Zn] vs [Zn/H] for our data as well as from the literature. Again, only detections have been plotted.
 Both figures 11 and 12 show a trend toward higher depletion with higher metallicity. 
 Previous investigations have shown a similar anti-correlation between [Cr/Zn] or [Fe/Zn] and [Zn/H] 
although the vast majority of their data set has come from absorbers at $z > 1.5$ \citep{Pet97, Pro02, Ak05}.
  Our study increases the number of Zn measurements at $z < 1.0$ by $\sim$ 30$\%$
  A Spearman rank-order correlation test on this data set gives R$_{\rm S}$=-0.428, with a probability of obtaining such a value by
 chance of P(R$_{\rm S}$)=6.00$\times$10$^{-3}$. 

\begin{figure}
\vspace{8mm}
\includegraphics[width=\linewidth]{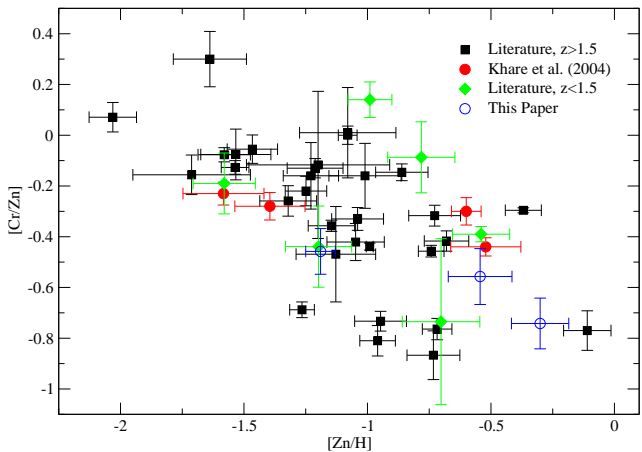}
\vspace{2mm}
\caption{[Cr/Zn] vs. [Zn/H] for our data as well as measurements from the literature. Only detections are plotted.}
\end{figure}

\begin{figure}
\vspace{4mm}
\includegraphics[width=\linewidth]{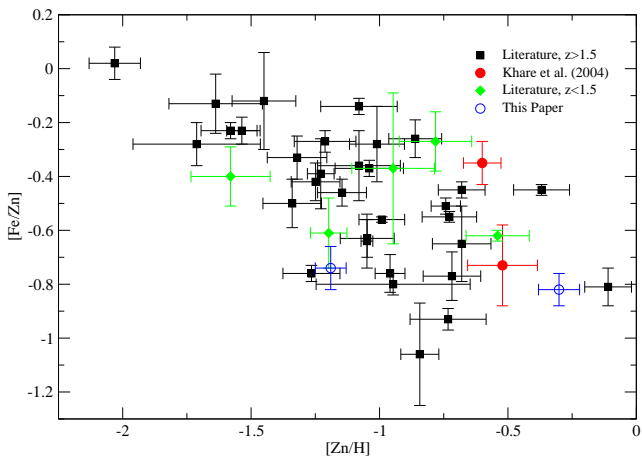}
\vspace{2mm}
\caption{[Fe/Zn] vs. [Zn/H] for our data as well as from the literature. Only detections are plotted.}
\end{figure}

\subsection{Dust Obscuration Bias}    

Several studies have tried to understand whether current DLA observations are affected by a dust obscuration bias. 
  Recently, York et al. (2006) conducted a survey of 809 Mg II 
 systems from the SDSS with 1.0 $\leq$ $z$ $\leq$ 1.9 and found that QSOs with an intervening
 absorber are at least three times more
 likely to have highly reddened spectra than QSOs without any absorption systems in their spectra. Furthermore, York et al. (2006) found a
 trend toward higher Zn II column densities but lower [Zn/H] and lower abundances of the first ions as the reddening E(B-V) increased. 
Significant reddening has also been seen in a sample of strong Ca II absorbers \citep{WH05,Wild06}.

 If dust obscuration effects are significant, highly reddened QSOs would be under-sampled due to their faintness. 
 Boisse et al. (1998) pointed out that there is a deficit of systems with high $N_{\rm H I}$ and high [Zn/H], and 
suggested that this may be caused by a dust obscuration bias (because such systems may also have high dust content and 
may obscure the background quasars more). They speculated that observations of fainter QSOs might reveal systems 
with high $N_{\rm H I}$ and high 
metallicity. Figure 13 shows a plot of [Zn/H] vs. $N_{\rm H I}$ for DLAs from our sample as well as the literature 
(Akerman et al. 2005, Rao et al. 2005). Clearly, there is a deficit of systems that are seen with both low
 metallicity and low $N_{\rm H I}$. This could be attributed to observational limitations in detecting weak Zn II lines, 
 although recent high resolution, high S/N observations still find a dearth of systems in this region. More
 interestingly though, there are
 few absorbers with high metallicity and high $N_{\rm H I}$, as noted by Boisse et al. (1998) and also by Akerman et al. (2005). 
 This cannot be credited to observational limitations, because systems
 with similar metallicities have been detected at lower $N_{\rm H I}$, and seems to suggest that these types of systems are being
 under-sampled.  Also, there is nothing unphysical
 about systems with high $N_{\rm H I}$ and high metallicity. It is thus surprising that only 
 3 points from the literature (of which 1 came from our previous 
 sample in Khare et al. 2004) lie above the dashed line indicating the 
 empirical ``obscuration threshold" log $N_{\rm Zn II}$ $<$ 13.15 suggested by 
 Boisse et al. (1998).

  Two of the systems at $z < 1.5$ that lie above the ``obscuration threshold'' came from our samples (this paper and Khare et al. 2004), 
 although it is still remarkable that all points in Fig. 13 lie fairly close to the threshold suggested by Boisse et al. (1998). 
Indeed, all but two of the points are consistent, within the errors, with being below the obscuration threshold. It should also be 
noted that the Boisse et al. threshold is not meant to be a hard limit, but an observationally determined boundary based on the 
much smaller sample available to Boisse et al. (1998). Indeed, hydrodynamical simulations by Cen et al. (2003) predict 
the existence of absorbers above this threshold. Thus, it may not be completely surprising that we are now finding some 
objects above the threshold.

We emphasize that the higher proportion of systems with N$_{\rm Zn II}$ close to the Boisse et al. threshold 
in our studies (this paper and Khare et al. 2004) is not due to errors in the column density estimations from the 
moderate resolution of the MMT spectrograph. Column densities derived from our observations of several systems with 
both the MMT and with the higher resolution 
 ($\sim$ 5 km s$^{-1}$) Ultraviolet-Visual Echelle Spectrograph (UVES) on the Very Large Telescope (VLT) agree
 to within 0.1 dex \citep{Pe06a, Pe06b}. A similar agreement was also found between column densities derived 
by Khare et al. (2004) from MMT spectra and those derived by Rao et al. (2005b) from Keck HIRES spectra. The reason for 
the higher proportion of metal-rich systems 
in our studies is that the targets in this
 paper and Khare et al. (2004) were chosen partially because of strong metal line absorption features seen in SDSS spectra. Thus, these
 systems may have been more likely to have higher Zn column densities. 
Indeed, composite SDSS quasar spectra indicate that systems with
larger W$_{rest}$ of Mg II $\lambda 2796$ tend to have stronger Zn II lines
(Nestor et al. 2003; York et al. 2006).
 One of our two systems has a large depletion consistent with the obscuration selection effect while the other system
 shows a moderate depletion, so it is not clear 
whether the small number of such systems found so far is caused by dust obscuration or small sample size especially at $z
 < 1.5$ where the effects of dust are expected to be the most pronounced \citep{Boi98}. It could also be the case that the most
  metal rich systems are truly rare. 

Figures 14 and 15 show the dust content as measured by [Cr/Zn] and [Fe/Zn] respectively vs. log $N_{\rm H I}$ for 
 these DLAs as well as those from the literature. As can be seen, there is very little trend in the data.
It may appear surprising from Figs. 11-15 that there is no (anti)-correlation between
[Cr/Zn] (or [Fe/Zn]) and log N(HI) even though there is an (anti)-correlation between [Cr/Zn]
(or [Fe/Zn]) and [Zn/H] and between [Zn/H] and log N$_{\rm HI}$. If one traces some of the
extreme points in Figs. 11 (or 12) and 13 that do show the correlation, one can see
that the correlations in these two figures arise from different systems.
For example, consider the 3 points A, B, and C with ([Cr/Zn], [Zn/H], log N$_{\rm HI}$ $\approx$ 
(-0.46, -1.19, 21.52); (-0.74, -0.30, 21.10); and (-0.54; -.57; 20.85).
Points such as A and B are responsible for the correlation in Fig 11.
 Points such as B and C are responsible
for the correlation in Fig. 13. When one plots these points in Fig. 14, there is no
strong correlation. (More generally, if data sets x,y and y,z show correlations,
x and z do not necessarily show a correlation; whether or not they do
depends on the distribution of the individual data values. See, for example, Casella $\&$ Berger 2002.)

\subsection{Constraints on Metallicities and Potential Implications for Metallicity Evolution} 

The Zn abundances for our DLAs are listed in Table 5.  In 4 of the systems at 
$0.6 < z < 1.5$ the abundances of Zn
 are about 10$\%$ solar or lower, while in 2 systems, the abundances are 30-50 $\%$ solar. We now briefly discuss the implications of our
 data for the metallicity evolution of DLAs. Our analysis is based on the statistical
 procedures outlined in Kulkarni \& Fall (2002), and uses our data, the data 
 compiled in Kulkarni et al. (2005), and other recent data
 from the literature (Rao et al. 2005; Akerman et al. 2005).
 We binned the  combined sample of 109 DLAs in 6 redshift bins with 18 or 19
 systems each and calculated the global N$_{\rm H I}$-weighted metallicity in 
 each bin. Figure 16 shows [Zn/H] vs. $z$  
 for the data in the literature as well as from this paper.

 We constructed two samples for these 109 DLAs: a
 ``maximum limits'' sample where the Zn limits are treated as detections, and 
 a ``minimum limits'' sample, where the Zn limits are treated
 as zeros. For an individual system these extreme cases cover the full range of
  possible values the Zn column densities can take in the
 case of the limits. The $N_{\rm H I}$-weighted mean metallicity in the lowest 
 redshift bin $0.1 < z < 1.2$ is $- 0.78 \pm
 0.11$ for the ``maximum limits'' sample and
 $- 0.99 \pm 0.17$ for the ``minimum limits'' sample. The linear regression
 slope of the metallicity-redshift relation for the redshift range 
 $0.1 < z < 3.9$ is $-0.18 \pm 0.07$ for the
 ``maximum limits'' sample, and $-0.25 \pm 0.10$ for the ``minimum limits'' 
 sample. The corresponding estimates for the intercept of the
 metallicity-redshift relation are $-0.65 \pm 0.15$ for the ``maximum limits'' 
 sample and $-0.63 \pm 0.21$ for the ``minimum limits''
 sample. Thus our data support the conclusions of Khare et al. (2004) and 
 Kulkarni et al. (2005) that the global mean metallicity of DLAs
 shows at best a slow evolution with redshift.

\section{Conclusions and Future Work}

The MMT observations presented here, together with our previous MMT and HST data 
have more than doubled the DLA Zn sample at $z < 1.5$ and more than tripled the sample at $z < 1$. 
Combining our data with data from the literature, we find that 
 the systems with higher [Zn/H] also have stronger Cr depletion [Cr/Zn]. 
  Analysis of the $N_{\rm H I}$-weighted mean metallicity vs. redshift for 
 our sample combined with previous data from the literature supports previous conclusions 
 that the $N_{\rm H I}$-weighted mean global DLA metallicity rises slowly at best and 
 does not reach solar levels by $z = 0$. 

  Questions still remain pertaining to dust depletion and selection effects, 
  and linked to this, the metallicity-redshift relationship of
 DLAs. Despite the large improvement from our surveys, DLA samples at $z < 1.5$ 
 are still relatively small. Clearly, more observations of DLAs 
 at $z < 1.5$ are needed for improved statistics in this vast epoch ($\sim 9$ Gyr). 
 The large number of SDSS absorbers now available present a good opportunity 
 for this purpose.

\begin{figure}
\includegraphics[width=\linewidth]{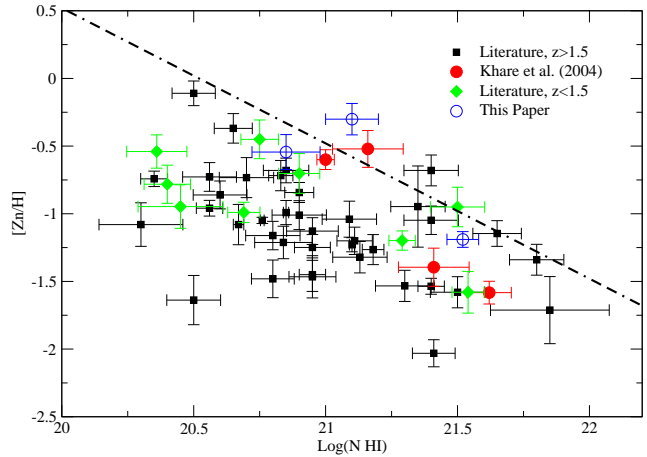}
\vspace{2mm}
\caption{[Zn/H] vs. log $N_{\rm H I}$ for our data as well as previous 
measurements from the literature. The dashed line indicates the empirical 
upper limit log $N_{\rm Zn II} < 13.15$ inferred from previous studies 
and suspected to be the ``obscuration threshold''. Only detections are plotted.} 
\end{figure}

\begin{figure}
\vspace{7mm}
\includegraphics[width=\linewidth]{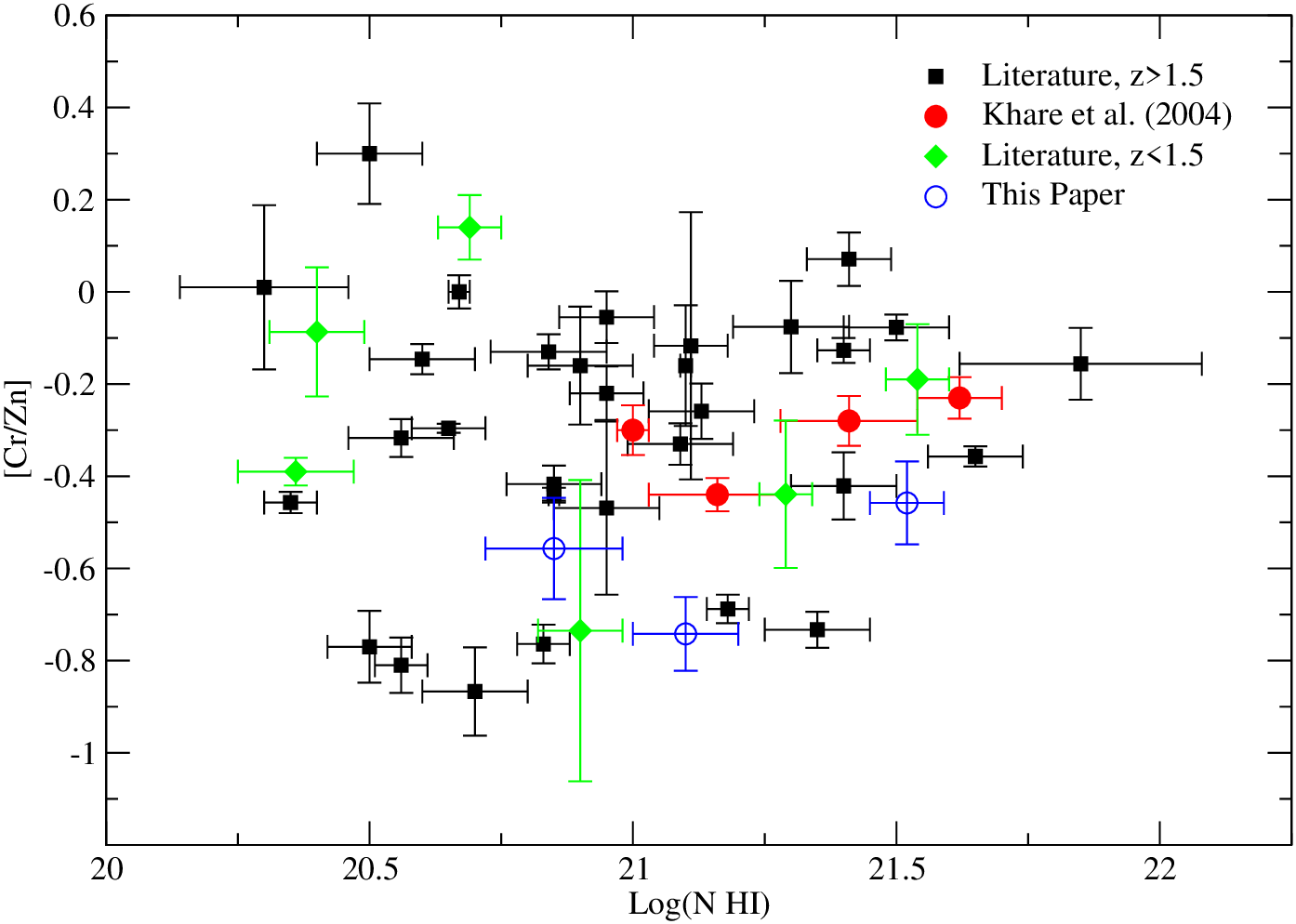}
\vspace{4mm}
\caption{[Cr/Zn] vs. log $N_{\rm H I}$ for our data as well as measurements 
from the literature. Only detections are plotted.}

\end{figure}

\begin{figure}
\vspace{5mm}
\includegraphics[width=\linewidth]{fig15.eps}
\vspace{3mm}
\caption{[Fe/Zn] vs. log $N_{\rm H I}$ for our data as well as measurements 
from the literature. Only detections are plotted.} 

\end{figure}

\begin{figure}
\vspace{7mm}
\includegraphics[width=\linewidth]{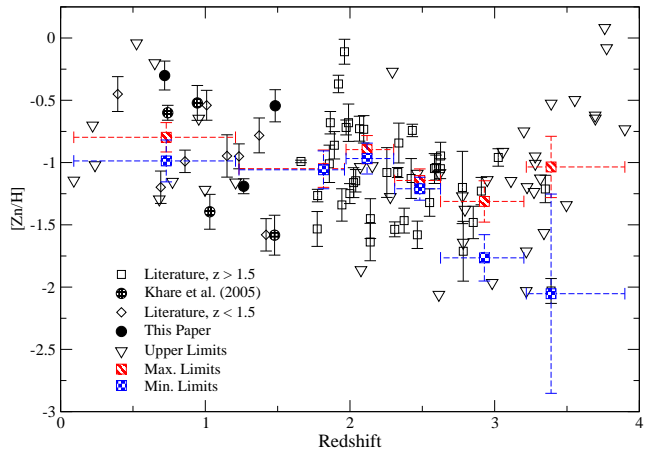}
\vspace{4mm}
\caption{The metallicity [Zn/H] vs. z for the DLAs in this sample as well as from the literature.} 
\end{figure}

 Two of our systems have higher $N_{\rm Zn II}$ than the empirical upper limit (log $N_{\rm Zn II} < 13.15$) that has been 
 attributed to dust obscuration in previous studies. The fairly high Zn abundances 
 in  these 2 DLAs suggests that metal-rich DLAs may be a rare
 class of objects, but can be found in a large enough sample. It would be 
 interesting to confirm the metal-rich nature of these DLAs with
 higher resolution spectra in the future, and find similar other systems by 
 observing a larger sample. If a large number of such systems are found, 
 they may make a significant contribution to the cosmic budget of metals. 
 
 It has been suggested that the low metallicity found in DLAs is caused by metallicity gradients. On the basis of a small sample
 of objects Chen et al. (2005) reported a difference between emission line and absorption line metallicities. If metallicity gradients are
 present in DLAs, then metal poor DLAs may be more likely to be found because larger impact parameters are more probable. 
 This may be because (a) they correspond to larger cross-sections and (b) they correspond to less dust. On the other hand, other studies
 \citep{Sch05, Bow05} have found no difference between emission and absorption line metallicities for other DLAs. Given that the
 metallicity gradients are not well
 established and fairly weak even in nearby spirals, it is not clear if metallicity gradients fully explain the low metallicities seen in
 DLAs.

 So far, no DLA has been observed with a super-solar metallicity. The key
 to finding such systems may lie in observing faint, reddened QSOs. Vladilo \& P\'eroux (2005) have estimated that dusty DLAs may 
 contribute as much as 17 $\%$ of the total metals. 
 $\Delta$($g$ - $i$) measurements are available from SDSS photometry for thousands of quasars with 
 candidate DLAs and sub-DLAs. It would be interesting to obtain [Cr/Zn] and [Fe/Zn] for a 
 large sample of such objects to study dust depletion patterns in quasar
 absorbers and to examine trends between quasar reddening and dust depletion.

\section*{Acknowledgments}
 We thank the staff of the MMT observatory for assistance 
 during our observing runs. We thank C. P\'eroux 
 for helpful comments on an earlier version of this paper and J. Grego for discussions on statistical techniques.
 JM and VPK gratefully acknowledge support from the National Science 
 Foundation grant AST-0206197 and the NASA/STScI grant GO-9441. PK acknowledges 
 support from the University of South Carolina Research Foundation during 
 a visit there in Spring and Summer 2004. We also would like to thank the anonymous referee
 for the insightful and helpful comments concerning this paper.

\bsp

\label{lastpage}

\end{document}